\documentstyle[12pt]{article}
\makeatletter
\@addtoreset{equation}{section}

\makeatother

\begin{document}

\begin{titlepage}
\null
\begin{flushright}
UT-929
\\
hep-th/0104208
\\
April, 2001
\end{flushright}

\vskip 1.5cm
\begin{center}


{\Large Coset Construction of $Spin(7), G_2$ Gravitational
 Instantons}

\lineskip .75em
\vskip 2.5cm

\normalsize

  {Y. Konishi
\footnote{email: konishi@hep-th.phys.s.u-tokyo.ac.jp}
 and M. Naka
\footnote{email: naka@hep-th.phys.s.u-tokyo.ac.jp}}

\vskip 1.5em

  {\it Department of Physics, University of Tokyo\\
               Tokyo 113-0033, Japan}

\vskip 2cm

{\bf Abstract}
\end{center}

We study Ricci-flat metrics on non-compact manifolds
with the exceptional holonomy $Spin(7), G_2$. 
We concentrate on the metrics which are defined on ${\bf R} \times G/H$.
If the homogeneous coset spaces $G/H$ have weak $G_2$,
$SU(3)$ holonomy,
the manifold ${\bf R} \times G/H$ may have $Spin(7), G_2$ holonomy metrics.        
Using the formulation with vector fields,
we investigate the metrics with $Spin(7)$ holonomy
on ${\bf R}\times Sp(2)/Sp(1), {\bf R}\times SU(3)/U(1)$.
We have found the explicit volume-preserving vector fields
on these manifold using the elementary coordinate parameterization.
This construction is essentially dual to solving the generalized
self-duality condition for spin connections.
We present most general differential equations
for each coset.
Then, we develop the similar formulation in order to 
calculate metrics with $G_2$ holonomy.

\baselineskip=0.6cm

\clearpage

\end{titlepage}

\section{Introduction}
\hspace{5mm}
It is important to search explicit metrics with
exceptional holonomy in order to
discuss the string compactifications with the lower amount of
supersymmetry.
Recently, 
Cveti\v{c}, Gibbons, L\"u and Pope 
have obtained new complete Ricci-flat
metrics with $Spin(7)$ holonomy on bundles over ${\bf S}^4, {\bf CP}^2$ 
\cite{CvGiLuPo}.
These metrics can be regarded as the generalization of explicit metrics
of ${\bf R}^4$ bundle over ${\bf S}^4$ and
${\bf R}^4/{\bf Z}_2$ bundle over ${\bf CP}^2$
obtained in \cite{BrSa, GiPaPo} to Taub-NUT and Taub-BOLT metrics.
These metrics are constructed on ${\bf R}\times Sp(2)/Sp(1),
{\bf R}\times SU(3)/U(1)$ where
${\bf S}^7=Sp(2)/Sp(1)\to Sp(2)/(Sp(1)\times Sp(1))={\bf S}^4,
SU(3)/U(1)\to SU(3)/S(U(2)\times U(1))={\bf CP}^2$ are the
principal $SU(2), SU(2)/{\bf Z}_2$-bundles, respectively.
All these examples are metrics defined on ${\bf R} \times G/H$
where $G/H$ denotes the coset space.      
Here, we should address the classification of Ricci-flat solutions with
coset spaces.
For this problem, we wish to note that there seems to be no other
examples due to the results
on the classification of the compactification for eleven-dimensional
supergravity \cite{CaRoWa}.          
Sufficient condition can be given (for 
literature in physics, see \cite{AcFiHuSp}).
Conical spaces with base G/H have $Spin(7), G_2$
holonomy if $G/H$ have weak $G_2, SU(3)$
holonomy.
Such $G/H$ is classified in weak $G_2$ holonomy:
locally $Sp(2)/Sp(1), SU(3)/U(1)$.
Weak $SU(3)$ is not classified, but known examples are
$G_2/SU(3)$,
twistor spaces $Sp(2)/(Sp(1)\times U(1)),
SU(3)/(U(1)\times U(1))$ and
$SU(2)\times SU(2)$.

On the other hand, Yasui and Ootsuka \cite{YaOo} proposed 
a formulation using volume-preserving vector field of 
$Spin(7)$ holonomy metrics on
eight-dimensional manifolds, which is the analogue of the Ashtekar
gravity construction of HyperK\"ahler metrics in four-dimensions
\cite{AsJaSm, MaNe, Donaldson, HaYaMiOo}.
In this article, we apply this formulation to the spaces
${\bf R}\times Sp(2)/Sp(1), {\bf R}\times SU(3)/U(1)$. 
Then we obtain the same first order equations in \cite{CvGiLuPo} 
from which Ricci-flat solutions follow as
the certain condition using the vector fields naturally associated to
the coset space. 
Especially, we determine the explicit form of the volume-preserving
vector fields which plays the central role in the formulation.
Moreover, we give most general first order equations
for such construction on $SU(3)/U(1)$.
It would be interesting to find a fruitful application of the results. 
We also develop the similar criterion in order to find
the metric with $G_2$ holonomy by using vector fields.
We apply this  ${\bf R}\times {\bf S}^3\times {\bf S}^3$ found
in \cite{BrSa, GiPaPo}.
However, we have not found the volume-preserving vector fields
on ${\bf R}\times Sp(2)/(Sp(1)\times U(1)), {\bf R}\times
SU(3)/(U(1)\times U(1))$ in \cite{BrSa, GiPaPo}.
We have no definite answer whether such volume-preserving vector
fields should exist on these spaces.
The coset $G_2/SU(3) \simeq SO(7)/SO(6)$ does not give non-trivial solutions
other than flat seven-dimensional space.

There are certain sufficient conditions called the self-duality
conditions on the spin connection for metrics on eight-,
seven-dimensional manifolds to have $Spin(7), G_2$ holonomy, respectively
\cite{AcOL}.
In our coset examples, the self-duality conditions are equivalent to
the conditions for vector fields.
This is because both vector fields and dual one-forms needed
to obtain the spin connection are essentially determined by the Lie
algebra of the Lie group from which the coset space is constructed.

The organization of this paper is the following.
In section 2, we 
discuss coset spaces which have weak $G_2, SU(3)$
holonomy.
Then we review the known Ricci-flat metrics with exceptional
holonomy including that found in \cite{CvGiLuPo}.
In section 3, 
we study the condition for $Spin(7)$ holonomy metrics 
by following \cite{YaOo}.
We write down the volume-preserving vector fields.
We find more general differential equations given by \cite{CvGiLuPo}.
In section 4, we introduce a similar criterion for $G_2$ holonomy metrics on
seven-manifolds and apply it to the metrics ${\bf R}\times {\bf S}^3\times {\bf
S}^3$ discovered in \cite{BrSa, GiPaPo}.
We briefly discuss why the formulation do not seem to work for the
remaining two examples on
${\bf R}\times Sp(2)/(Sp(1)\times U(1)), {\bf R}\times
SU(3)/(U(1)\times U(1))$ in \cite{BrSa, GiPaPo}.
We include section 5 for the conclusions and discussions.
Two appendices provide various notations necessary for this article. 

\section{Spaces ${\bf R} \times G/H$ with Exceptional Holonomy}
\hspace{5mm}

At first, let us remind the definition of the exceptional holonomy metrics.
We introduce the concept of specific differential forms on it.
Then, we discuss weak holonomy spaces and coset spaces.

\vspace{3mm}

A metric with $Spin(7)$ holonomy on the simply connected
eight-dimensional manifold $M$
admits a 
$Spin(7)$ invariant covariantly-constant spinor $\eta$.
If such $\eta$ exists, it is automatically Ricci-flat.
Such metric has the $Spin(7)$ invariant, self-dual, and closed
four-form $\Omega$ (called Cayley four-form).
The components can be written down as $\Omega_{\alpha\beta\gamma\delta}=
\bar{\eta}\Gamma_{\alpha\beta\gamma\delta}\eta$.
By the suitable choice of the orthonormal basis $\{e^{\alpha}\}$
$(\alpha=1,\dots,8)$,
the Cayley four-form is given by the following form
\begin{equation}
\label{Cayley}
\Omega=\frac{1}{4!}\Psi_{\alpha\beta\gamma\delta}
e^{\alpha} \wedge e^{\beta} \wedge e^{\gamma} \wedge e^{\delta}.
\end{equation}
The $\Psi_{\alpha\beta\gamma\delta}$ are defined by the
structure constants of octonion algebra
\begin{equation}
\label{comp}
\Psi_{abc8}=c_{abc},\qquad \Psi_{abcd}=\frac{1}{3!}
\epsilon_{abcd}^{\quad\;\; efg}c_{efg},
\end{equation}
where $a=1,\dots,7$.
We use the representation that the non-zero components of 
totally antisymmetric tensor $c_{abc}$ are
\begin{equation}
\label{oct}
c_{123}=c_{516}=c_{624}=c_{435}=c_{471}=c_{673}=c_{572}=1. 
\end{equation}
Then, the explicit expression of the Cayley four-form $\Omega$ is
\begin{eqnarray}
\label{cayley}
\Omega&=& e^1\wedge e^2 \wedge e^3 \wedge e^8 +
        e^5\wedge e^1 \wedge e^6 \wedge e^8 +
         e^6\wedge e^2 \wedge e^4 \wedge e^8 +
        e^4\wedge e^3 \wedge e^5 \wedge e^8\nonumber\\ 
       && + e^4\wedge e^7 \wedge e^1 \wedge e^8 
         + e^6\wedge e^7 \wedge e^3 \wedge e^8 +
        e^5\wedge e^7 \wedge e^2 \wedge e^8 \nonumber\\
       &&+e^4\wedge e^5 \wedge e^6 \wedge e^7 +
        e^2\wedge e^3 \wedge e^7 \wedge e^4 +
         e^1\wedge e^3 \wedge e^5 \wedge e^7 +
        e^1\wedge e^2 \wedge e^7 \wedge e^6\nonumber\\ 
 && + e^2\wedge e^3 \wedge e^5 \wedge e^6 
         + e^1\wedge e^2 \wedge e^4 \wedge e^5 +
        e^1\wedge e^3 \wedge e^4 \wedge e^6. 
\end{eqnarray}
Conversely, if the four-form $\Omega$ (\ref{Cayley}) 
with respect to a
metric on $M$ is closed, 
then the metric has $Spin(7)$ holonomy \cite{BrSa}.

Here, we address the relation to the self-duality equation of
the spin connection.
In general, the spin connection 
$\omega_{\alpha\beta}$ takes the value
in the Lie algebra of $SO(8)$.
In \cite{AcOL}, the condition of which the spin connection on  
$Spin(7)$ holonomy manifold should satisfy is proposed:
\begin{equation}
\label{spin}
\omega_{\alpha\beta}=\frac{1}{2}\Psi_{\alpha\beta}^{\quad\gamma\delta}
\omega_{\gamma\delta}.
\end{equation}
This is the generalized self-duality condition in eight-dimension.
We may call this equation as the octonionic instanton equation.
Recall that, in four-dimension,
the self-duality equation for the spin-connection is based on the 
structure constants of the quaternionic algebra.
In fact, we can show that if the spin connection satisfies the
self-duality condition (\ref{spin}), the four-form (\ref{cayley}) is closed:
\begin{equation}
d\Omega=0.
\end{equation}
Using this property, certain metric on ${\bf R}^4$ bundles over 
${\bf S}^4$ in \cite{BrSa, GiPaPo} 
was rederived as the octonionic gravitational
instanton \cite{BaFlKe}.

A metric on a simply-connected seven-dimensional manifold has $G_2$
holonomy if and only if the three form $\Phi$ below and its dual
four-form $*\Phi$ are closed.
Here, by the suitable choice of the orthonormal basis, $\Phi$
can be written as
\begin{equation}
\label{g2form}
\Phi=\frac{1}{3!}c_{abc}e^a\wedge e^b \wedge e^c,
\end{equation}
where $c_{abc}$ are structure constants of octonionic algebra
 (\ref{oct}).

In \cite{AcOL},  
the self-duality condition for the spin connections on manifolds with
$G_2$ holonomy was written as follows 
\begin{equation}
\omega_{ab}=\frac{1}{2}\Psi_{ab}^{\quad cd }\omega_{cd},
\end{equation}
where $\Psi_{abcd}$ is the same as (\ref{Cayley}).
If the spin connection satisfies this self-duality
condition, the forms (\ref{g2form}) are closed
\begin{equation}
d\Phi=d*\Phi=0.
\end{equation}

\vspace{3mm}

In search of the metrics with exceptional holonomy groups, 
we will look into the $d$-dimensional space which is locally of the form 
${\bf R}\times M_{d-1}$ and $M_{d-1}$ is a level surface for each $t$
(the holonomy group is $Spin(7)$ for $d=8$ and $G_2$ for $d=6$ ). 
Then what is the necessary and sufficient condition for $M_{d-1}$ 
for ${\bf R}\times {M_{d-1}}$ to admit the metric with exceptional holonomy ? 
The sufficient condition is known 
that ${\bf R}\times M_7$ admits the metric with $Spin(7)$ holonomy 
if $M_7$ admits the metric 
with  weak $G_2$-holonomy 
and that ${\bf R}\times M_6$ admit the metric with 
$G_2$-holonomy 
if $M_6$ admits the metric 
with weak $SU(3)$-holonomy. 

First we discuss the case of $Spin(7)$-holonomy ($d=8$). 
The metric of a simply-connected 8-dimensional manifold has  $Spin(7)$-holonomy
if and only if the self-dual four-form $\Omega$
(2.1) is closed. 
And the metric of a 7-dimensional manifold $M_7$ is said to have 
weak $G_2$-holonomy if and only if the three-form $\Phi$ (2.7) satisfies 
\begin{equation}
d\Phi=\lambda *\Phi
\end{equation}
where $\lambda$ is a constant \cite{Gray}.
We will consider the metric 
\begin{equation}
dt^2+ds^2_{M_7}(t)
\end{equation}
on ${\bf R}\times M_7$ 
which becomes one parameter family of metrics on $M_7$ when 
restricted to $t=constant$. Then the three form $\Phi(t)$ 
(2.7) with respect to
the metric $ds^2_{M_7}(t)$ and its dual $*\Psi(t)$ give the four-form (2.1) 
\begin{equation}
\Omega=dt\wedge \Phi(t)+*\Phi(t).
\end{equation}
The closure of $\Omega$ is equivalent to the 
evolution equation \cite{Hitchin}
\begin{equation}
{{\partial}\over{\partial t}}*\Phi(t)=d\Phi(t).
\end{equation}
If $M_7$ admits the metric with weak $G_2$-holonomy, 
the existence of the solution is assured in its neighborhood:
let $\Phi$ be the three form (2.7) with respect to the metric with 
weak $G_2$-holonomy and assume at $t=t_0$ $\Phi(t)=\Phi$. 
Then the evolution equation at $t=t_0$ is
\begin{equation}
{{\partial }\over{\partial t}}*\Phi(t)=\lambda *\Phi
\end{equation}
and the following solution exists, 
\begin{equation}
*\Phi(t)\sim (1+\lambda (t-t_0))*\Phi
\end{equation}
for small $t-t_0$.

Next we will discuss the case of $G_2$-holonomy ($d=7$). 
The metric of a simply-connected 7-dimensional manifold has  $G_2$-holonomy
if and only if both the three-form $\Phi$
(2.7) and its dual four-form $*\Phi$ are closed \cite{BrSa}. 
And the metric of a 6-dimensional manifold $M_6$ is said to have 
weak $SU(3)$-holonomy if and only if the three-form
\begin{equation}
\rho=e_1\wedge e_2\wedge e_3-e_3\wedge e_4\wedge e_5
+e_1\wedge e_3\wedge e_6-e_4\wedge e_2\wedge e_6
\end{equation}
 and the two-form 
\begin{equation}
\omega=e_5\wedge e_6+e_1\wedge e_4+e_3\wedge e_2
\end{equation}
satisfy
\begin{equation}
d\hat{\rho}=-2\lambda *\omega^2,
\hskip 0.2 in
d\omega=3\lambda\rho 
\hskip 0.2 in
(\lambda:\mbox{constant}) 
\end{equation}
where $\{e_i\}$is the orthonormal frame and $\hat{\rho}$ is determined by 
the property that $\rho+i\hat{\rho}$ is a complex (3,0)-form preserved by 
$SL(3,C)$ \cite{Hitchin}, specifically
\begin{equation}
\hat{\rho}=e_3\wedge e_4\wedge e_6-e_1\wedge e_2\wedge e_6
+e_1\wedge e_3\wedge e_5-e_4\wedge e_2\wedge e_5.
\end{equation}
As in the $spin(7)$ case, we will consider the metric 
\begin{equation}
dt^2+ds^2_{M_6}(t)
\end{equation}
on ${\bf R}\times M_6$ which becomes one parameter 
family of metrics on $M_6$ when 
restricted to $t=constant$. 
Then the three-form $\rho(t)$ and 
the two-form $\omega(t)$ with respect to
the metric $ds^2_{M_6}(t)$ give the three-form (2.7) 
\begin{equation}
\Phi=dt\wedge \omega(t)+\rho(t).
\end{equation}
The closure of $\Phi$ is equivalent to the 
Hamiltonian flow equation \cite{Hitchin}
\begin{equation}
{{\partial}\over{\partial t}}\rho(t)=d\omega(t),
\hskip 0.3 in 
{{\partial}\over{\partial t }}\omega^2=-2d \hat{\rho}.
\end{equation}
If $M_6$ admits the metric with weak $SU(3)$-holonomy, 
the existence of the solution is assured in its neighborhood:
let $\rho$ be the three form and $\omega$  the two-form 
with respect to the metric with weak 
$SU(3)$-holonomy and assume at $t=t_0$ 
$\rho(t_0)=\rho, \omega(t_0)=\omega$. 
Then the evolution equation at $t=t_0$ is
\begin{equation}
{{\partial}\over{\partial t}}\rho(t)=3\lambda\rho,
\hskip 0.3 in 
{{\partial}\over{\partial t }}\omega^2=4\lambda\omega^2
\end{equation}
and the following solution exists, 
\begin{equation}
\rho(t)\sim (1+3\lambda (t-t_0))\rho,
\hskip 0.3 in
\omega^2(t)\sim(1+4\lambda (t-t_0))\omega^2
\end{equation}
for small $t-t_0$.

\vspace{3mm}

We restrict the level surface $M_{d-1}$ to homogeneous coset spaces. 
The cosets which give the metrics with $Spin(7)$
holonomy on ${\bf R} \times G/H$ 
was classified in relation to the compactification of
the eleven-dimensional supergravity \cite{CaRoWa}.
Let us consider the metric of M2 brane configuration
which is placed on conical singularity of the cone $C_8$
\begin{equation}
ds^2=H(r)^{-\frac{2}{3}}ds_3^2+H(r)^{\frac{1}{3}}\left(
dr^2+r^2ds_{G/H}^2\right),
\end{equation}
where $H(r)=1+\left(\frac{a}{r}\right)^6$
is a harmonic function on $C_8$.
The metric approaches to that on 
the product space of three-dimensional Minkowski space and
the cone at $r \to \infty$.
Around $r \to 0$, we can rewrite this metric into
near horizon geometry $AdS_4 \times G/H$ with $r=\sqrt{2au}$.
These three supergravity backgrounds are
(1) M2 brane configuration (2.25),
(2) $M_3 \times C_8$,
(3) $AdS_4 \times G/H$.
Except for round sphere, the number of supersymmetry 
on (1), (2) is half of that on (3) \cite{AcFiHuSp}.
The number of supersymmetry on (2)
is determined by the number of parallel spinors on the cone.
If the cone is simply-connected, the number of parallel spinors  
can be read off from the irreducible holonomy of the cone \cite{Bar}.
Thus, we can read the number of supersymmetries on brane configuration
from the holonomy of the cone.
Such type of consideration naturally fits the idea of holography 
\cite{AcFiHuSp}.    
It is known that cone with level surface $M_7$
has holonomy contained in $Spin(7)$ if $M_7$ have weak $G_2$ holonomy.
Except for the discrete quotient on $G/H$,
coset spaces with weak $G_2$ holonomy are known to be
\begin{equation}
Sp(2)/Sp(1), \qquad SU(3)/U(1).
\end{equation}
The list in \cite{CaRoWa, AcFiHuSp} also includes $SO(5)/SO(3)$.
This is locally equivalent to $Sp(2)/Sp(1)$.

As for $G_2$ holonomy manifolds, 
supergravity vacua containing $M_6$
with maximal supersymmetry are not known.
However, again it is known that
cone with level surface $M_6$ has $G_2$
holonomy if and only if $M_6$ has weak $SU(3)$ holonomy.
The known coset spaces with weak $SU(3)$ holonomy are
\begin{equation}
Sp(2)/(Sp(1)\times U(1))
\quad SU(3)/(U(1)\times U(1)) \quad 
SU(2)\times SU(2) \quad G_2/SU(3)
\end{equation}
It seems that coset $G_2/SU(3)$ does not permit 
explicit $G_2$ holonomy metric.
It only permits the metric on ${\bf R}^7$.

We do not know the necessary and sufficient condition on $M_{d-1}$
to obtain $Spin(7), G_2$ holonomy spaces ${\bf R} \times M_{d-1}$.
Thus, in this paper, we restrict to the coset spaces (2.26) and (2.27).
These cosets include the all known examples with analytic solution.

\vspace{3mm}

In the rest of this section, 
we collect the ansatz and the first order differential
equations from which the known Ricci-flat solutions can be derived.
These equations are derived in \cite{CvGiLuPo,CvGiLuPo2} using 
a calculation of Ricci tensors and the superpotentials.
In \cite{CvGiLuPo, CvGiLuPo2}, it was shown that
these first order equations are the integrability conditions for the 
existence of a single covariantly constant spinor in the eight- or
seven-dimensional metrics.
Thus, the solutions for the first order equations should always give the
metrics with special holonomy.
More general first order equations will derived in section 3.

\hspace{2mm}

{\it $Spin(7)$ holonomy}

\hspace{2mm}

The ansatz can be written as follows \cite{CvGiLuPo}:
\begin{equation}\label{spin7}
ds^2=dt^2+A(t)^2(\sigma_1^2+\sigma_2^2)
+B(t)^2\sigma_3^2+C(t)^2 \left\{
\begin{array}{l}
{{1}\over{2}}d\Omega_4^2\\
d\Sigma_2^2\end{array}
\right.  ,
\end{equation}
where  
$SU(2)$ is described as a $U(1)$ bundle over ${\bf S}^2$ ;
\begin{eqnarray}
\label{fib1}
i\sigma_1+\sigma_2&=&
-{{e^{-i\psi}}\over{2}}(d\theta+i\sin\theta d\phi)\\
&&\; +
e^{-i\psi}
\left(i\cos\theta\;\Im[e^{-i\phi}(i{\cal A}_1+{\cal A}_2)]
+\sin\theta\;\Re[e^{-i\phi}(i{\cal A}_1+{\cal A}_2)]
-i\sin\theta {\cal A}_3
\right),\nonumber\\
\sigma_3&=&{{1}\over{2}}(d\psi+\cos\theta d\phi)
+\cos\theta{\cal A}_3+
\sin\theta\;\Im[e^{-i\phi}(i{\cal A}_1+{\cal A}_2)],
\end{eqnarray}
where the periods of $(\theta,\psi,\phi)$ are $(2\pi, 4\pi, 2\pi)$, 
$(2\pi,2\pi,2\pi)$ for ${\bf S}^4$, ${\bf CP}^2$, respectively. 
$d\Omega_4^2, d\Sigma_2^2$ is respectively
the standard round metric on ${\bf S}^4$, 
the Fubini-Study metric on ${\bf CP}^2$, 
with cosmological constant
$\Lambda=6, 3$ and 
${\cal A}_1,{\cal A}_2,{\cal A}_3$ 
is the self-dual $SU(2)$-connection over ${\bf S}^{4}$ (\ref{s4instanton}) or 
${\bf CP}^2$ (\ref{p2instanton}). 
The corresponding field strength $F^i$
satisfy the quaternionic algebra and can be regarded as almost complex
structures \cite{GiPaPo}. 

Then, the first order equations from which the Ricci-flat metrics with
$Spin(7)$ holonomy follow are
\begin{equation}
\label{spin7}
\dot{A}=2-\frac{B}{A}-\frac{A^2}{C^2},\quad
\dot{B}=\frac{B^2}{A^2}-\frac{B^2}{C^2},\quad
\dot{C}=\frac{A}{C}+\frac{B}{2C},
\end{equation}
where the dot $\dot{}$ represents a derivative with respect to $t$. 
These differential equations were solved in \cite{CvGiLuPo}. 

Here, we include the following remarks.
\begin{itemize}
\item
We should note that these metrics are defined on
the following fibration structures
\begin{equation}
\label{coset}
\frac{Sp(2)}{Sp(1)}\to \frac{Sp(2)}{Sp(1)\times Sp(1)}
, \qquad \frac{SU(3)}{U(1)} \to \frac{SU(3)}{S(U(2)\times U(1))}.
\end{equation}

\item
From the coset structures (\ref{coset}), 
we can generalize the ansatz of the metric by allowing 
the coefficients of $\sigma_1$ and $\sigma_2$ to be different:
$A^2(\sigma_1^2+\sigma_2^2)\rightarrow 
A_1^2\sigma_1^2+A_2^2\sigma_2^2$.
The differential equation for this ansatz is (\ref{difeq1}).
\end{itemize}

\hspace{2mm}

{\it $G_2$ holonomy}

\hspace{2mm}

The ansatz for ${\bf R}^4$ bundle over ${\bf S}^3$
can be written as follows \cite{BrSa, GiPaPo}:
\begin{equation}
ds^2=dt^2+A(t)^2\left(\Sigma^i-\frac{1}{2}\sigma^i\right)^2
+C(t)^2\sigma^i\sigma^i,
\end{equation}
where $\Sigma^i, \sigma^i$ are the left-invariant one-forms on the group
manifold $SU(2)$.
Here, these one-forms are normalized by 
$d\Sigma^i=2\epsilon_{ijk}\Sigma^j\wedge\Sigma^k,\;
d\sigma^i=2\epsilon_{ijk}\sigma^j\wedge\sigma^k$
(This normalization is different from the appendix in \cite{CvGiLuPo2}).
It is easily shown that the $G_2$ structure is invariant under
the $SU(2)$ action of the fiber ${\bf S}^3$.
Then, the first order equations from which the Ricci-flat metrics with
$G_2$ holonomy follows are
\begin{equation}\label{g21}
\dot{A}=1-\frac{A^2}{4C^2},\quad
\dot{C}=\frac{A}{2C}.
\end{equation}

\hspace{2mm}

The ansatz for ${\bf R}^3$ bundle over ${\bf S}^4, {\bf CP}^2$
can be written down as follows \cite{BrSa, GiPaPo}:
\begin{equation}
\label{g22}
ds^2=dt^2+A(t)^2(\sigma_1^2+\sigma_2^2)
+C(t)^2\left\{
\begin{array}{l}
{{1}\over{2}}d\Omega_4^2\\
d\Sigma_2^2\end{array}
\right.
\end{equation}
where 
$\sigma_i$'s are (\ref{fib1}) in the $Spin(7)$ case with $\psi$ set to
be zero. 
$d\Omega_4^2$ and $d\Sigma_2^2$ are the standard metric on ${\bf S}^4$,
the Fubini-Study metric on ${\bf CP}^2$.
Then, the first order equations from which the Ricci-flat metrics with
$G_2$ holonomy follow are
\begin{equation}\label{g2eq}
\dot{A}=2-\frac{A^2}{C^2},\quad
\dot{C}=\frac{A}{C}.
\end{equation}

\hspace{2mm}

We can observe that these metrics are based on the following fibrations
\begin{equation}
SU(2) \times SU(2) \to SU(2),
\end{equation}
\begin{equation}
\frac{Sp(2)}{Sp(1)\times U(1)}\to \frac{Sp(2)}{Sp(1)\times Sp(1)}
, \frac{SU(3)}{U(1)\times U(1)} \to \frac{SU(3)}{S(U(2)\times U(1))}.
\end{equation}

We could show that the first order equations (\ref{g21})
is derived by using above self-duality condition.
However, it is difficult to obtain
first order equations (\ref{g22}) by using self-duality equations 
for the remaining two metrics on
${\bf R}^3$ bundle over ${\bf S}^4, {\bf CP}^2$.

\section{$Spin(7)$ Holonomy Metrics}
\hspace{5mm}

In \cite{YaOo}, another construction of metrics with $Spin(7)$
holonomy was introduced.
The metrics are constructed from 
the solutions to the first-order differential
equations for the volume-preserving vector fields.
We summarize the results in the following.

\hspace{2mm}

{\it Proposition} \cite{YaOo}

\hspace{2mm}

Let $M$ be a simply-connected eight-dimensional manifold, and $\omega$
a volume form on $M$.
We introduce the linearly independent vector fields 
$V_{\alpha}\:(\alpha=1,\dots,8)$ on $M$.
We denote the one-forms dual to $V_{\alpha}$ by $W^{\alpha}$.
Suppose that the vector fields $V_{\alpha}$
satisfy the following two conditions:

(1) volume-preserving condition

\begin{equation}
{\cal L}_{V_{\alpha}}\omega=0.
\end{equation}

(2) 2-vector condition

\begin{equation}
\label{vector}
\Psi_{\alpha\beta\gamma\delta}\;
[V_{\alpha}\wedge V_{\beta}, V_{\gamma}\wedge V_{\delta}]_{{\rm SN}}=0,
\end{equation}
where $\Psi_{\alpha\beta\gamma\delta}$ is the same as (\ref{comp}).
And $[\:,\:]_{{\rm SN}}$ is the Schouten-Nijenhuis bracket, i.e.
\begin{eqnarray}
[V_{\alpha}\wedge V_{\beta}, V_{\gamma}\wedge V_{\delta}]_{{\rm SN}}
&=&
[V_{\alpha}, V_{\gamma}]\wedge V_{\beta} \wedge V_{\delta}\\
&&-
[V_{\alpha}, V_{\delta}]\wedge V_{\beta} \wedge V_{\gamma}
-
[V_{\beta}, V_{\gamma}]\wedge V_{\alpha} \wedge V_{\delta}
+
[V_{\beta}, V_{\delta}]\wedge V_{\alpha} \wedge V_{\gamma}.
\nonumber
\end{eqnarray}
Then the metric with $Spin(7)$ holonomy can be written down as follows
\begin{equation}
g=\phi\; W^{\alpha} \otimes W^{\alpha},
\end{equation}
where $\phi^2=\omega(V_1,V_2,\dots, V_8)$.
The corresponding Cayley four-form is given by
\begin{equation}
\Omega=\frac{1}{4!}\;\phi^2 \;\Omega_{\alpha\beta\gamma\delta}
\; W^{\alpha}\wedge W^{\beta} \wedge W^{\gamma} \wedge W^{\delta}.
\end{equation}

\hspace{2mm}

{\it Application} 

\hspace{2mm}

We apply the above formulation to the following two eight-dimensional
spaces :
\begin{equation}
M={\bf R} \times \frac{Sp(2)}{Sp(1)}, \quad
{\bf R} \times \frac{SU(3)}{U(1)},
\end{equation}
which are the total spaces of coset spaces (\ref{coset}).
Detailed notations are summarized in the appendix.

\hspace{2mm}

${\bf R}\times {Sp(2)/Sp(1)}$

\hspace{2mm}

The coset space $Sp(2)/Sp(1)$ is a $SU(2)$ principal bundle over ${\bf S}^4$.
The coordinates are $(x_0,x_1,x_2,x_3,\theta,\psi,\phi)$ 
where $x_i$ are the coordinates on ${\bf S}^4$ and 
$(\theta,\psi,\phi)$ are those on $SU(2)$. 
$(\theta,\psi,\phi)$ have period $(2\pi, 4\pi, 2\pi)$, respectively. 
We denote $x_0^2+x_1^2+x_2^2+x_3^2$ by $|x|^2$.

Let $\sigma_1, \dots, \sigma_7$ be local one-forms on
$Sp(2)/Sp(1)$  which are left-invariant one-forms
when pulled back to $Sp(2)$. They are 
\begin{eqnarray}
i\sigma_1+\sigma_2&=&e^{-i\psi}\Biggl[
-{{1}\over{2}}\left(d\theta+i\sin\theta d\phi\right)
\nonumber\\
&&\;+
i\cos\theta\;\Im[e^{-i\phi}(i{\cal A}_1+{\cal A}_2)]
+\sin\theta\;\Re[e^{-i\phi}(i{\cal A}_1+{\cal A}_2)]
-i\sin\theta{\cal A}_3
\Biggr],
\nonumber\\
\sigma_3&=&{{1}\over{2}}(d\psi+\cos\theta d\phi)
+\cos\theta{\cal A}_3+
\sin\theta\;\Im[e^{-i\phi}(i{\cal A}_1+{\cal A}_2)],
\\
\left(
\begin{array}{c}
i\sigma_4+\sigma_5\\i\sigma_6+\sigma_7
\end{array}
\right)&=&{{\sqrt{2}}\over{1+|x|^2}}
\left(
\begin{array}{cc}
\cos{{\theta}\over{2}}e^{ -i(\psi+\phi)/2}&
\sin{{\theta}\over{2}}e^{ -i(\psi-\phi)/2}\\
-\sin{{\theta}\over{2}}e^{i(\psi+\phi)/2}&
\cos{{\theta}\over{2}}e^{i(\psi-\phi)/2}
\end{array}
\right)
\left(\begin{array}{c}
-idx_1+dx_2\\idx_3+dx_0
\end{array}\right),\nonumber
\label{sigmasp}
\end{eqnarray}
where the $SU(2)$ connections ${\cal A}_i$ are 
\begin{eqnarray}
{\cal A}_1={{1}\over{1+|x|^2}}
\left( 
x_2dx_0-x_0dx_2+x_3dx_1-x_1dx_3
\right),\nonumber\\
{\cal A}_2={{1}\over{1+|x|^2}}
\left( 
x_0dx_1-x_1dx_0+x_3dx_2-x_2dx_3
\right),\nonumber\\\label{s4instanton}
{\cal A}_3={{1}\over{1+|x|^2}}
\left( 
x_0dx_3-x_3dx_0+x_2dx_1-x_1dx_2
\right).
\end{eqnarray}
The differentials of the left-invariant one-forms are 
\begin{eqnarray}
d{\sigma}_1&=&
2{\sigma}_2\wedge{\sigma}_3+{\sigma}_4\wedge{\sigma}_7
-{\sigma}_5\wedge{\sigma}_6,
\nonumber
\\
d{\sigma}_2&=&
-2{\sigma}_1\wedge{\sigma}_3+{\sigma}_4\wedge{\sigma}_6
+{\sigma}_5\wedge{\sigma}_7,
\nonumber
\\
d{\sigma}_3&=&
2{\sigma}_1\wedge{\sigma}_2+{\sigma}_4\wedge{\sigma}_5
-{\sigma}_6\wedge{\sigma}_7,
\nonumber
\\
d{\sigma}_4&=&
-{\sigma}_1\wedge{\sigma}_7-{\sigma}_2\wedge{\sigma}_6
-{\sigma}_3\wedge{\sigma}_5+{\sigma}_5\wedge{\sigma}_{10}
+{\sigma}_6\wedge{\sigma}_9-{\sigma}_7\wedge{\sigma}_8,
\nonumber
\\
d{\sigma}_5&=&
{\sigma}_1\wedge{\sigma}_6+{\sigma}_3\wedge{\sigma}_4
-{\sigma}_2\wedge{\sigma}_7-{\sigma}_4\wedge{\sigma}_{10}
-{\sigma}_6\wedge{\sigma}_8-{\sigma}_7\wedge{\sigma}_9,
\nonumber
\\
d{\sigma}_6&=&
-{\sigma}_1\wedge{\sigma}_5+{\sigma}_2\wedge{\sigma}_4
+{\sigma}_3\wedge{\sigma}_7+{\sigma}_5\wedge{\sigma}_8
-{\sigma}_4\wedge{\sigma}_9+{\sigma}_7\wedge{\sigma}_{10},
\nonumber
\\
\label{dif11}
d{\sigma}_7&=&
{\sigma}_1\wedge{\sigma}_4+{\sigma}_2\wedge{\sigma}_5
-{\sigma}_3\wedge{\sigma}_6+{\sigma}_4\wedge{\sigma}_8
+{\sigma}_5\wedge{\sigma}_9-{\sigma}_6\wedge{\sigma}_{10},
\end{eqnarray}
where 
\begin{equation}
\sigma_8=\sum_{i=4}^7A^8_i\sigma_i,\quad
\sigma_9=\sum_{i=4}^7A^9_i\sigma_i,\quad
\sigma_{10}=\sum_{i=4}^7A^{10}_i\sigma_i.
\end{equation}
The functions $A_j^8,A_j^9,A_j^{10}$ are obtained in the Appendix B, 
but all that is needed here is  the identity 
\begin{eqnarray}
-{{3}\over{2}}\;{{d|x|^2}\over{1+|x|^2}}&=& 
(A_7^8-A_6^9-A_5^{10})\sigma_4+
(A^8_6+A_7^9+A^{10}_4)\sigma_5
\nonumber\\
\label{identity1}
&&\;+(-A_5^8+A_4^9-A_7^{10})\sigma_6+
(-A^8_4-A^9_5+A_6^{10})\sigma_7.
\end{eqnarray}
The detailed derivation of (\ref{sigmasp}) $-$ (\ref{identity1}) is 
presented in the Appendix B.

Equipped with these coset structures, 
first we look for the local volume form and  local volume preserving
vector-fields on $Sp(2)/Sp(1)$. Our ansatz is 
\begin{eqnarray}\nonumber
\mbox{volume form}&=&F\sigma_1\wedge\sigma_2\wedge\sigma_3\wedge\sigma_4
\wedge\sigma_5\wedge\sigma_6\wedge\sigma_7,\\
\mbox{vector fields}&=&f_1X_1,\cdots, f_7X_7,
\end{eqnarray}
where $X_1,\cdots, X_7$ are dual vector fields of $\sigma_1,\cdots, \sigma_7$
and $F, f_i$ are local functions on coset space $Sp(2)/Sp(1)$.
Using the formula ${\cal L}_{X_i}=d\iota_{X_i}+\iota_{X_i}d$ , 
the volume preserving condition 
${\cal L}_{f_iX_i}(F\sigma_1\wedge\cdots)=0$
turned out to be
\begin{eqnarray}
d_{X_1}(Ff_1)=d_{X_2}(Ff_2)=d_{X_3}(Ff_3)=0,
\nonumber
\\
d_{X_4}(Ff_4)+(Ff_4)(A^8_7-A^9_6-A^{10}_5)=0,
\nonumber
\\
d_{X_5}(Ff_5)+(Ff_4)(A^8_6+A^9_7+A^{10}_4)=0,
\nonumber
\\
d_{X_6}(Ff_6)+(Ff_6)(-A^8_5+A^9_4-A^{10}_7)=0,
\nonumber
\\
d_{X_7}(Ff_7)+(Ff_7)(-A^8_4-A^9_5+A^{10}_6)=0.
\end{eqnarray}
By the above identity (\ref{identity1}), we can find a solution:
\begin{eqnarray}
Ff_1, Ff_2, Ff_3 &=& \mbox{functions of }x,
\nonumber
\\\label{vpc1}
Ff_4=Ff_5=Ff_6=Ff_7&=&(1+|x|^2)^{3/2}\quad
\mbox{up to multiplication of constants.}
\end{eqnarray}

Next we apply the proposition \cite{YaOo}. 
Let $r$ be a coordinate on ${\bf R}$.
We put the ansatz for the volume form and the 
vector fields on ${\bf R}\times Sp(2)/Sp(1)$ as follows:
\begin{eqnarray}
\nonumber
\mbox{volume form}&=&F\sigma_1\wedge\sigma_2\wedge\sigma_3\wedge\sigma_4
\wedge\sigma_5\wedge\sigma_6\wedge\sigma_7\wedge dr, \\
\label{vfield}
\mbox{vector fields}&:&
V_i=a_i(r)f_iX_i \hskip 0.1 in (1\leq i\leq 7),\quad
V_8=f_8{{\partial}\over{\partial r}},
\end{eqnarray}
where $F,f_i (1\leq i\leq 8)$ are functions on $Sp(2)/Sp(1)$.
These vector fields preserves the volume form when (\ref{vpc1}) is 
satisfied. If they also satisfy the 2-vector condition
 (\ref{vector}), 
which will be described later, 
the following metric has $Spin(7)$-holonomy:
\begin{equation}
ds^2=(Ff_1\cdots f_7f_8)^{1/2}\; (a_1(r)\cdots a_7(r))^{1/2}\times
\left(
{{(\sigma_1)^2}\over{f_1^2a_1^2}}+\cdots
+{{(\sigma_7)^2}\over{f_7^2a_7^2}}+
{{dr^2}\over{f_8^2}}
\right).
\end{equation}
The natural requirement that this metric  be well-defined 
not only locally but globally on $Sp(2)/Sp(1)$
determines $F$ and $f_i$: 
\begin{eqnarray}
{{(Ff_1\cdots f_8)^{1/2}}\over{f_i^2}}=1,
\nonumber\\
a_4(r)^2=a_5(r)^2=a_6(r)^2=a_7(r)^2.
\end{eqnarray}
\begin{equation}\label{fspin7}
\Rightarrow f=(1+|x|^2)^{-1/2},\quad F=f^{-4},
\end{equation}
where $f=f_1=\dots=f_8$.

Here, we need the  
Lie brackets of $X_i$'s 
to compute the 2-vector condition (\ref{vector}). 
They are obtained by using (\ref{liebrackets}) 
without knowing the explicit form of $X_i$'s:
$$
[X_1,X_2]=-2X_3,\quad
[X_1,X_3]=2X_2,\quad
[X_1,X_4]=-X_7,
$$$$
[X_1,X_5]=X_6,\quad
[X_1,X_6]=-X_5,\quad
[X_1,X_7]=X_4,
$$$$
[X_2,X_3]=-2X_1,\quad
[X_2,X_4]=-X_6,\quad
[X_2,X_5]=-X_7,
$$$$
[X_2,X_6]=X_4,\quad
[X_2,X_7]=X_5,\quad
[X_3,X_4]=-X_5,
$$$$
[X_3,X_5]=X_4,\quad
[X_3,X_6]=X_7,\quad
[X_3,X_7]=-X_6,
$$$$
[X_4,X_5]=-X_3+A^{10}_4X_4+A^{10}_5X_5
          +(A_4^8+A_5^9)X_6+(-A_5^8+A_4^9)X_7,
$$$$
[X_4,X_6]=-X_2+A^9_4X_4+(-A_4^8+A_6^{10})X_5
          +A^9_6X_6+(-A_6^8-A_4^{10})X_7,
$$$$
[X_4,X_7]=-X_1-A_4^8X_4+(-A_4^9+A_7^{10})X_5
          +(A_7^9+A_4^{10})X_6-A_7^8X_7,
$$$$
[X_5,X_6]=X_1+(A_5^9-A_6^{10})X_4+A_5^8X_5
          -A_6^8X_6+(-A_6^9-A_5^{10})X_7,
$$$$
[X_5,X_7]=-X_2+(-A_5^8-A_7^{10})X_4-A_5^9X_5
          +(-A_7^8-A_5^{10})X_6-A_7^9X_7,
$$
\begin{equation}\label{xixjsp}
[X_6,X_7]=X_3+(-A_6^8-A^9_7)X_4+(A_7^8-A_6^9)X_5
          +A_6^{10}X_6+A_7^{10}X_7.
\end{equation}
If we substitute the ansatz (\ref{vfield}), (\ref{fspin7}) into 
the 2-vector condition (\ref{vector}), 
the miraculous cancellation occurs and the 2-vector condition becomes as 
follows
\begin{eqnarray}
\label{first}
{{a_1^{\prime}}\over{a_1}}+{{a_2^{\prime}}\over{a_2}}+
{{b^{\prime}}\over{b}}-
{{2c^2}\over{a_1}}-{{2c^2}\over{a_2}}+
{{2c^2}\over{b}}=0,&&
{{b^{\prime}}\over{b}}+{{2c^{\prime}}\over{c}}+
{{c^2}\over{b}}-{{2a_1a_2}\over{b}}-2a_1-2a_2=0,
\nonumber\\
{{a_1^{\prime}}\over{a_1}}+{{2c^{\prime}}\over{c}}
-{{c^2}\over{a_1}}-2a_2+2b-2b{{a_2}\over{a_1}}=0,&&
{{a_2^{\prime}}\over{a_2}}+{{2c^{\prime}}\over{c}}
-{{c^2}\over{a_2}}-2a_1+2b-2b{{a_1}\over{a_2}}=0,
\end{eqnarray}
where we put $b=a_3, c=a_4=a_5=a_6=a_7$.
The parameter $|x|$ in vector fields disappears in these differential
equations.

For the solutions to (\ref{first}), we have the metrics with $Spin(7)$
holonomy
\begin{equation}
\label{metric}
ds^2=\sqrt{a_1a_2b}c^2\left(
dr^2+\frac{1}{a_1^2}\sigma_1^2+\frac{1}{a_2^2}\sigma_2^2
+\frac{1}{b^2}\sigma_3^2
+\frac{1}{c^2}(\sigma_4^2+\sigma_5^2+
\sigma_6^2+\sigma_7^2)
\right).
\end{equation} 
Note that  
$\sigma_4^2+\sigma_5^2+\sigma_6^2+\sigma_7^2$ is the 
half of the round metric on ${\bf S}^4$.
In order to see the relation to the solutions in \cite{CvGiLuPo}, 
we rewrite above metric into the following form
\begin{equation}
\label{spin72}
ds^2=
dt^2+A_1(t)^2\sigma_1^2+A_2(t)^2\sigma_2^2
+B(t)^2\sigma_3^2
+C(t)^2(\sigma_4^2+\sigma_5^2+
\sigma_6^2+\sigma_7^2).
\end{equation}
Then, changing the coordinate from $r$ to $ t$ by 
$(a_1a_2b)^{1/4}cdr=dt$, 
we obtain the following first-order differential equations
\begin{eqnarray}
\label{difeq1}
\nonumber
\dot{A_1}&=&2+\frac{B}{A_2}-\frac{A_1^2}{C^2}
+\frac{A_1}{B}\left(\frac{A_2}{A_1}-\frac{A_1}{A_2}\right),
\quad
\dot{A_2}=2+\frac{B}{A_1}-\frac{A_2^2}{C^2}
-\frac{A_2}{B}\left(\frac{A_2}{A_1}-\frac{A_1}{A_2}\right),
\\
\dot{B}&=&\frac{B^2}{C^2}-\frac{B^2}{A_1A_2}
+{{A_2}\over{A_1}}+{{A_1}\over{A_2}}-2,
\qquad
\dot{C}=\frac{A_1+A_2}{2C}-\frac{B}{2C},
\end{eqnarray}
where the dot $\dot{}$ denotes the differentiation by $t$.
With the identification $B\rightarrow -B$ and setting $A_1=A_2$, 
we have the same first-order differential equations as (\ref{spin7}).
The Cayley four-form for this metric is (\ref{Cayley}) with 
the orthonarmal basis
\begin{eqnarray}
\label{basis}
e_1=A_1\sigma_1,\quad
e_2=A_2\sigma_2,\quad
e_3=B\sigma_3,\nonumber\\
e_4=C\sigma_4,\quad
e_5=C\sigma_5,\quad
e_6=C\sigma_6,\quad
e_7=C\sigma_7,\nonumber\\
e_8=dt.
\end{eqnarray}

Another way of obtaining the metric with $Spin(7)$ holonomy (\ref{spin72}) is 
to start with the ansatz of the orthonormal basis
 (\ref{basis}) and impose the closure of the Cayley 
four form (\ref{Cayley}). Then the differential equation
 (\ref{difeq1}) is derived. 

\hspace{2mm}

$SU(3)/U(1)$

\hspace{2mm}

This case is almost same as the case of $Sp(2)/Sp(1)$.
The coset space $SU(3)/U(1)$ is a principal bundle over ${\bf CP}^2$
with fiber $SU(2)/{\bf Z}_2$. 
On this space, we introduce the coordinates $(w_1, \overline{w}_1, w_2, \overline{w}_2,
\theta, \psi, \phi)$ 
where $(w_1, w_2)$ are the inhomogeneous complex 
coordinates of ${\bf CP}^2$ and $(\theta, \psi, \phi)$ are 
those on $SU(2)$.
The Periods of $(\theta, \psi, \phi)$ are $(2\pi, 2\pi, 2\pi)$.

Let $\sigma_1, \dots, \sigma_7$ be local one-forms on
$SU(3)/U(1)$  which are left-invariant one-forms
when pulled back to $SU(3)$. They are 
\begin{eqnarray}
\label{cp2form}
i\sigma_1+\sigma_2&=&
e^{-i\psi}\Biggl[-{{1}\over{2}}\left(d\theta+i\sin\theta d\phi-i\sin\theta{\cal A}_3)
\right.\nonumber\\
&&\; + 
i\cos\theta\;\Im[e^{-i\phi}(i{\cal A}_1+{\cal A}_2)]
+\sin\theta\;\Re[e^{-i\phi}(i{\cal A}_1+{\cal A}_2)]
\Biggr],\\
\nonumber
\sigma_3&=&{{1}\over{2}}(d\psi+\cos\theta d\phi)
+\cos\theta{\cal A}_3+
\sin\theta\;\Im[e^{-i\phi}(i{\cal A}_1+{\cal A}_2)],
\\
\left(
\begin{array}{c}
i{\sigma}_4+{\sigma}_5\\
i{\sigma}_6+{\sigma}_7
\end{array}
\right)&=&
\left(\begin{array}{cc}
\cos{{\theta}\over{2}}e^{-i(\psi+\phi)/2}&
\sin{{\theta}\over{2}}e^{-i(\psi-\phi)/2}\\
-\sin{{\theta}\over{2}}e^{i(\psi-\phi)/2}&
\cos{{\theta}\over{2}}e^{i(\psi+\phi)/2}
\end{array}
\right)
\left(\begin{array}{c}
{{dw_1}\over{\sqrt{(1+|w_1|^2)(1+|w|^2)}}}\\
{{-\overline{w}_1w_2dw_1+(1+|w_1|^2)dw_2}
\over{(1+|w|^2)\sqrt{1+|w_1|^2}}}
\end{array}
\right),\nonumber
\end{eqnarray}
where ${\cal A}_i$ are
\begin{eqnarray}
i{\cal A}_1+{\cal A}_2&=&
{{-\overline{w}_2dw_1}\over{(1+|w_1|^2)\sqrt{1+|w|^2}}}
,\nonumber\\\label{p2instanton}
{\cal A}_3&=&{{1}\over{2i}}\left(
-{{\overline{w}_1d{w}_1}\over{1+|w_1|^2}}+
{{\overline{w}_1dw_1+\overline{w}_2dw_2}\over{2(1+|w|^2)}}-c.c.
\right)
.\end{eqnarray}
Here, $|w|^2$ denotes $|w_1|^2+|w_2|^2$.

The differential of  $\sigma_i$'s are computed from the structure 
constant of $su(3)$:
\begin{eqnarray}
d\sigma_1&=&
2\sigma_2\wedge\sigma_3+\sigma_4\wedge\sigma_7
-\sigma_5\wedge\sigma_6,
\nonumber\\
d\sigma_2&=&
-2\sigma_1\wedge\sigma_3+\sigma_4\wedge\sigma_6
+\sigma_5\wedge\sigma_7,
\nonumber\\
d\sigma_3&=&
2\sigma_1\wedge\sigma_2+\sigma_4\wedge\sigma_5
-\sigma_6\wedge\sigma_7,
\nonumber\\
d\sigma_4&=&
-\sigma_1\wedge\sigma_7-\sigma_2\wedge\sigma_6
-\sigma_3\wedge\sigma_5+3\sigma_5\wedge\sigma_8,
\nonumber\\
d\sigma_5&=&
\sigma_1\wedge\sigma_6+\sigma_3\wedge\sigma_4
-\sigma_2\wedge\sigma_7-3\sigma_4\wedge\sigma_8,
\nonumber\\
d\sigma_6&=&
-\sigma_1\wedge\sigma_5+\sigma_2\wedge\sigma_4
+\sigma_3\wedge\sigma_7+3\sigma_7\wedge\sigma_8,
\nonumber\\
\label{dsigma2}
d\sigma_7&=&
\sigma_1\wedge\sigma_4+\sigma_2\wedge\sigma_5
-\sigma_3\wedge\sigma_6-3\sigma_6\wedge\sigma_8.
\end{eqnarray}
where 
\begin{equation}
\sigma_8=\sum_{i=4}^7A_i^8\sigma_i.
\end{equation}
We do not need the explicit form of the functions $A_i^8$. 
We have only to notice the identity they satisfy
\begin{equation}\label{identity2}
-{{1}\over{4}}{{d|w|^2}\over{1+|w|^2}}=
A_5^8\sigma_4-A_4^8\sigma_5+A_7^8\sigma_6-A_6^8\sigma_7.
\end{equation} 
More details of these calculation are presented in the Appendix B.

First we look for the volume form and  volume preserving
vector-fields on $SU(3)/U(1)$. Our ansatz is 
\begin{eqnarray}\nonumber
\mbox{volume form}&=&F\sigma_1\wedge\sigma_2\wedge\sigma_3\wedge\sigma_4
\wedge\sigma_5\wedge\sigma_6\wedge\sigma_7,\\
\mbox{vector fields}&=&f_1X_1, \cdots, f_7X_7,
\end{eqnarray}
where $X_1, \cdots, X_7$ are dual vector fields of $\sigma_1,\cdots,\sigma_7$
and $F,f_i$ are local functions on $SU(3)/U(1)$.
Using the formula ${\cal L}_{X_i}=d\iota_{X_i}+\iota_{X_i}d$ , 
the volume preserving condition ${\cal L}_{f_iX_i}
(F\sigma_1\cdots)=0$
becomes
\begin{eqnarray}\nonumber
d_{X_1}(Ff_1)=d_{X_2}(Ff_2)=d_{X_3}(Ff_3)=0,
\\\nonumber
d_{X_4}(Ff_4)+3(Ff_4)A^8_5=0,\hskip 0.2 in
d_{X_5}(Ff_5)-3(Ff_5)A^8_4=0,
\\
d_{X_6}(Ff_6)+3(Ff_6)A^8_7=0,\hskip 0.2 in
d_{X_7}(Ff_7)-3(Ff_7)A^8_6=0.
\end{eqnarray}
By the identity (\ref{identity2}), these partial differential 
equations can be solved:
$$
Ff_1, Ff_2, Ff_3 =\mbox{functions of }w,
$$\begin{equation}\label{vpc2}
Ff_4=Ff_5=Ff_6=Ff_7=(1+|w|^2)^{3/4}\hskip 0.1 in
\mbox{ up to multiplication of constants.}
\end{equation}

Next we apply the construction of \cite{YaOo} to  
${\bf R}\times SU(3)/U(1)$.
Let $r$ be a coordinate on ${\bf R}$.
We put the ansatz for the volume form and the 
vector fields on ${\bf R}\times SU(3)/U(1)$ as follows:
\begin{eqnarray}\nonumber
\mbox{vol. form}&=& F\sigma_1\wedge\sigma_2\wedge\sigma_3\wedge\sigma_4
\wedge\sigma_5\wedge\sigma_6\wedge\sigma_7\wedge dr, \nonumber\\\label{vfield2}
\mbox{vector fields}&:&
V_i=a_i(r)f_iX_i \hskip 0.1 in (1\leq i\leq 7),\quad
V_8=f_8{{\partial}\over{\partial r}},
\end{eqnarray}
where $F, f_i (1\leq i\leq 8)$ are functions on $SU(3)/U(1)$.
These vector fields preserves the volume form when (\ref{vpc2})
is satisfied. If they also satisfiy the 2-vector 
condition (\ref{vector}), 
which  we will  described later, 
the following metric has $Spin(7)$-holonomy:
\begin{equation}
ds^2=(Ff_1\cdots f_7f_8)^{1/2}(a_1(r)\cdots a_7(r))^{1/2}\times
\left(
{{(\sigma_1)^2}\over{f_1^2a_1^2}}+\cdots
{{(\sigma_7)^2}\over{f_7^2a_7^2}}+
{{dr^2}\over{f_8^2}}
\right)
\end{equation}
The further requirement that this metric  be well-defined 
not only locally but globally on $SU(3)/U(1)$
specifies $F$ and $f_i$: 
\begin{eqnarray}
{{(Ff_1\cdots f_8)^{1/2}}\over{f_i^2}}=1,
\nonumber\\
a_4(r)^2=a_5(r)^2,\quad a_6(r)^2=a_7(r)^2.
\end{eqnarray}
\begin{equation}
\Rightarrow f=(1+|w|^2)^{-1/4},\quad F=f^{-4}.
\end{equation}

Now, we need the Lie brackets of $X_i$'s 
to compute the 2-vector condition (\ref{vector}). 
They are obtained by using (\ref{liebrackets})  
even without knowing the explicit form of $X_i$'s:
$$
[X_1,X_2]=-2X_3,\hskip 0.2 in
[X_1,X_3]=2X_2,\hskip 0.2 in
[X_1,X_4]=-X_7,
$$$$
[X_1,X_5]=X_6,\hskip 0.2 in
[X_1,X_6]=-X_5,\hskip 0.2 in
[X_1,X_7]=X_4,
$$$$
[X_2,X_3]=-2X_1,\hskip 0.2 in
[X_2,X_4]=-X_6,\hskip 0.2 in
[X_2,X_5]=-X_7,
$$$$
[X_2,X_6]=X_4,\hskip 0.2 in
[X_2,X_7]=X_5,\hskip 0.2 in
[X_3,X_4]=-X_5,
$$$$
[X_3,X_5]=X_4,\hskip 0.2 in
[X_3,X_6]=X_7,\hskip 0.2 in
[X_3,X_7]=-X_6,
$$$$
[X_4,X_5]=-X_3+3A_4^8X_4+3A^8_5X_5,
$$$$
[X_4,X_6]=-X_2+3A_6^8X_5-3A_4^8X_7,
$$$$
[X_4,X_7]=-X_1+3A_7^8X_5+3A_4^8X_6,
$$$$
[X_5,X_6]=X_1-3A_6^8X_4-3A_5^8X_7,
$$$$
[X_5,X_7]=-X_2-3A_7^8X_4+3A_5^8X_6,
$$\begin{equation}\label{xixjsu}
[X_6,X_7]=X_3+3A_6^8X_6+3A_7^8X_7.
\end{equation}
We substitute the vector fields (\ref{vfield2}) into the 2-vector condition 
(\ref{vector}).
Then we obtain the differential equations
\begin{eqnarray}
\label{second}
{{a_1^{\prime}}\over{a_1}}+{{a_2^{\prime}}\over{a_2}}+
{{b^{\prime}}\over{b}}-
{{2c_1c_2}\over{a_1}}-{{2c_1c_2}\over{a_2}}+
{{c_1^2+c_2^2}\over{b}}=0,\nonumber
\\
{{b^{\prime}}\over{b}}+{{2c_1^{\prime}}\over{c_1}}+
{{c^2_2}\over{b}}-{{2a_1a_2}\over{b}}-2a_1-2a_2=0,\nonumber
\\
{{b^{\prime}}\over{b}}+{{2c_2^{\prime}}\over{c_2}}+
{{c^2_1}\over{b}}-{{2a_1a_2}\over{b}}-2a_1-2a_2=0,\nonumber
\\
{{a_1^{\prime}}\over{a_1}}+{{c_1^{\prime}}\over{c_1}}+{{c_2^{\prime}}\over{c_2}}
-{{c_1c_2}\over{a_1}}-2a_2+2b-2b{{a_2}\over{a_1}}=0,\nonumber
\\
{{a_2^{\prime}}\over{a_2}}+{{c_1^{\prime}}\over{c_1}}+{{c_2^{\prime}}\over{c_2}}
-{{c_1c_2}\over{a_2}}-2a_1+2b-2b{{a_1}\over{a_2}}=0,
\end{eqnarray}
where we put $b=a_3,c_1=a_4=a_5,c_2=a_6=a_7$.

For the solutions to (\ref{second}), we have the metrics with $Spin(7)$
holonomy
\begin{equation}
\label{metric2}
ds^2=\sqrt{a_1a_2b}c_1c_2\left(
dr^2+\frac{1}{a_1^2}\sigma_1^2+\frac{1}{a_2^2}\sigma_2^2
+\frac{1}{b^2}\sigma_3^2
+\frac{1}{c_1^2}(\sigma_4^2+\sigma_5^2)+\frac{1}{c_2^2}
(\sigma_6^2+\sigma_7^2)
\right).
\end{equation} 
Note that  if $c_1=c_2$, the differential equation is same as the case of 
$Sp(2)/Sp(1)$ and that the metric(\ref{metric2}) contains the 
Fubini-Study metric on ${\bf CP^2}$, 
$\sigma_4^2+\sigma_5^2+\sigma_6^2+\sigma_7^2$.
If we rewrite the metric on ${\bf R}\times SU(3)/U(1)$ 
above into the following form
\begin{equation}
\label{spin73}
ds^2=
dt^2+A_1(t)^2\sigma_1^2+A_2(t)^2\sigma_2^2
+B(t)^2\sigma_3^2
+C(t)^2_1(\sigma_4^2+\sigma_5^2)+
C(t)_2^2(\sigma_6^2+\sigma_7^2).
\end{equation}
by changing the coordinate from $r$ to $ t$ by 
$(a_1a_2b)^{1/4}(c_1c_2)^{1/2}dr=dt$, 
we obtain the following first-order differential equations
\begin{eqnarray}
\dot{A_1}=\frac{C_1^2+C_2^2}{C_1C_2}+\frac{B}{A_2}-\frac{A_1^2}{C_1C_2}
+\frac{A_1}{B}\left(\frac{A_2}{A_1}-\frac{A_1}{A_2}\right)
,\nonumber
\\
\dot{A_2}=\frac{C_1^2+C_2^2}{C_1C_2}+\frac{B}{A_1}-\frac{A_2^2}{C_1C_2}
-\frac{A_2}{B}\left(\frac{A_2}{A_1}-\frac{A_1}{A_2}\right),
\nonumber
\\
\label{difeq3}
\dot{B}=\frac{B^2}{2}\left(\frac{1}{C_1^2}+\frac{1}{C_2^2}\right)
-\frac{B^2}{A_1A_2}
+\frac{1}{B}\left(
{{A_2}\over{A_1}}+{{A_1}\over{A_2}}-2
\right),\nonumber
\\
\dot{C_1}=\frac{A_1+A_2}{2C_2}
-\frac{BC_1}{4}\left(\frac{1}{C_1^2}+\frac{1}{C_2^2}\right)
-\frac{A_3C_1}{4}\left(\frac{1}{C_1^2}-\frac{1}{C_2^2}\right)
-\left(\frac{1}{A_1}+\frac{1}{A_2}\right)\frac{C_1^2-C_2^2}{C_2},
\nonumber
\\
\dot{C_2}=\frac{A_1+A_2}{2C_2}
-\frac{BC_2}{4}\left(\frac{1}{C_1^2}+\frac{1}{C_2^2}\right)
-\frac{A_3C_2}{4}\left(\frac{1}{C_2^2}-\frac{1}{C_1^2}\right)
-\left(\frac{1}{A_1}+\frac{1}{A_2}\right)\frac{C_2^2-C_1^2}{C_1},
\end{eqnarray}
where $\dot{}$ denotes the differentiation by $t$.
The Cayley four-form for this metric is (\ref{Cayley}) with 
the orthonarmal basis
\begin{eqnarray}
e_1=A_1\sigma_1,\quad
e_2=A_2\sigma_2,\quad
e_3=B\sigma_3,\nonumber\\
e_4=C_1\sigma_4,\quad
e_5=C_1\sigma_5,\quad
e_6=C_2\sigma_6,\quad
e_7=C_2\sigma_7,\nonumber\\
e_8=dt.\label{basis}
\end{eqnarray}
As in  the case of $Sp(2)/Sp(1)$, the differential equation (\ref{spin73}) 
can also be obtained by imposing the closure of the Cayley four-form.

\section{$G_2$ Holonomy Metrics}
\hspace{5mm}

Now, we present the construction of metrics 
to the manifolds with $G_2$ holonomy.
We can summarize in the following proposition.

\hspace{3mm}

{\it Proposition}

\hspace{3mm}

Let $N$ be a simply-connected 
seven-dimensional manifold, and $\omega$ a volume form
on $N$.
We denote the linearly independent vector fields on $N$ by
$V_a\; (a=1,\dots,7)$.
And we define the one-forms dual to $V_a$ by $W^a$ such that
$W^a(V_b)=\delta_b^a$.
Let the vector fields $V_a$ and real functions $P,Q$ satisfy the following
conditions:

(1) volume-preserving condition
\begin{equation}  
{\cal L}_{V_a}\omega=0.
\end{equation}

(2) 
\begin{eqnarray}
&&\Phi_{abcd}\left( 2\frac{dP}{P}(V_a)\: V_b\wedge V_c \wedge V_d
+3[V_a, V_b] \wedge V_c \wedge V_d
\right)=0,\\
&&c_{abc}\left(
\frac{dQ}{Q}(V_a) \: V_b \wedge V_c +[V_a, V_b] \wedge V_c
\right)=0.
\end{eqnarray}

(3) 
\begin{equation}
\frac{Q^3}{P^4}=\omega(V_1, V_2,\dots,V_7).
\end{equation}

Then the metric with $G_2$ holonomy is given as follows
\begin{equation}
g=\left(\frac{Q}{P}\right)^2 W^a \otimes W^a.
\end{equation}
The corresponding three-form 
\begin{equation}
\Phi=\frac{1}{3!}\left(\frac{Q}{P}\right)^3\;
c_{abc}\;
W^a\wedge W^b \wedge W^c
\end{equation}
satisfies $d\Phi=d*\Phi=0$.

\hspace{3mm}

{\it Proof}

\hspace{3mm}

We assume that the metric is of the form 
\begin{equation}
g=\phi\; W^a \otimes W^a,
\end{equation}
where $\phi$ is a some real positive function.
In order for the metric to have $G_2$ holonomy, it is necessary and
sufficient to show that the following three-form
\begin{equation}
\Phi=\frac{1}{3!}\;
c_{abc}\;
\phi^{3/2}\;
W^a\wedge W^b \wedge W^c,
\end{equation}
satisfies $d\Phi=d*\Phi=0$.
Now we can rewrite
\begin{equation}
\Phi=\frac{\phi^{3/2}}{f}\frac{1}{4!}\;
\Phi_{abcd}\;
\iota_{V_a}\iota_{V_b}\iota_{V_c}\iota_{V_d}\omega,\quad
d*\Phi=\frac{\phi^2}{f}\frac{1}{3!}\;
c_{abc}\;
\iota_{V_a}\iota_{V_b}\iota_{V_c}\omega,
\end{equation}
where $f=\omega(V_1,\dots,V_8)$.
Then, using $d\omega={\cal L}_{V_a}\omega=0$, and 
\begin{equation}
{\cal L}_{V_a}\iota_{V_b}-\iota_{V_b}{\cal L}_{V_a}=\iota_{[V_a, V_b]},\quad
{\cal L}_{V_a}=d\iota_{V_a}+\iota_{V_a}d,
\end{equation}
\begin{equation}
\left(\beta\wedge\right)\iota_{V_\alpha}+\iota_{V_\alpha}\beta\wedge=
\beta(V_\alpha),
\end{equation}
where $\beta$ is a one-form and $\wedge$ means the external product,
we can show
\begin{eqnarray}
d\Phi=0 &\Leftrightarrow&
\Phi_{abcd}\left(
2\frac{d\left(\phi^{3/2}/f\right)(V_a)}{\phi^{3/2}/f}V_b\wedge V_c \wedge V_d
+3[V_a, V_b]\wedge V_c \wedge V_d
\right)=0,
\\
d*\Phi=0 &\Leftrightarrow&
c_{abc}\left(
\frac{d\left(\phi^2/f\right)(V_a)}{\phi^2/f} V_b\wedge V_c +[V_a, V_b] 
\wedge V_c \right)=0.
\end{eqnarray}
If we define $P=\phi^{3/2}/f, \:Q=\phi^2/f$, we obtain
\begin{equation}
f=\frac{Q^3}{P^4},\qquad \phi=\left(\frac{Q}{P}\right)^2.
\end{equation}
This completes the proof of the above proposition.

\hspace{3mm}

{\it Application}

\hspace{3mm}

We have succeeded to derive the first order equations on 
\begin{equation}
N= {\bf R} \times SU(2) \times SU(2).
\end{equation}

Let $\{\sigma_1,\sigma_2,\sigma_3\}$ and
$\{\Sigma_1,\Sigma_2,\Sigma_3\}$ be the left-invariant one-forms of the
first and the second $SU(2)$, respectively, and $r$ be a coordinate on
${\bf R}$.
We take $\sigma_1\wedge\sigma_2\wedge\sigma_3\wedge\Sigma_1
\wedge\Sigma_2\wedge\Sigma_3\wedge dr$ as the volume form
and use the following ansatz for the vector fields:
\begin{eqnarray}
\label{vec2}
&&V_1=c(r)\left(\sigma_1^*+\frac{1}{2}\Sigma_1^*\right),\quad
V_2=c(r)\left(\sigma_2^*+\frac{1}{2}\Sigma_2^*\right),\quad
V_3=c(r)\left(\sigma_3^*+\frac{1}{2}\Sigma_3^*\right),\nonumber\\
&&V_4=a(r)\Sigma_1^*,\quad
V_5=a(r)\Sigma_2^*,\quad
V_6=a(r)\Sigma_3^*,\nonumber\\
&&V_7=\frac{\partial}{\partial r}.
\end{eqnarray}
Corresponding dual one-forms are given by:
\begin{eqnarray}
&&W^1=\frac{1}{c(r)}\sigma_1,\quad
W^2=\frac{1}{c(r)}\sigma_2,\quad
W^3=\frac{1}{c(r)}\sigma_3,\nonumber\\
&&W^4=\frac{1}{a(r)}\left(\Sigma_1-\frac{1}{2}\sigma_1\right),\quad
W^5=\frac{1}{a(r)}\left(\Sigma_2-\frac{1}{2}\sigma_2\right),\quad
W^6=\frac{1}{a(r)}\left(\Sigma_3-\frac{1}{2}\sigma_3\right),\nonumber\\
&&W^7=dr.
\end{eqnarray}
Using the criterion (2), we have obtained the following first order
equations
\begin{equation}
\frac{Q'}{Q}+\frac{a'}{a}+\frac{c'}{c}-\frac{c^2}{2a}-2a=0,\quad
-\frac{P'}{P}=3\frac{a'}{a}-\frac{3c^2}{2a}=\frac{a'}{a}+2\frac{c'}{c}-2a.
\end{equation}
Then, combining with $\frac{Q^3}{P^4}=a^3c^3$, 
\begin{equation}
ds^2=dt^2+A(t)^2\left(\Sigma^i-\frac{1}{2}\sigma^i\right)^2
+C(t)^2\sigma^i\sigma^i,
\end{equation}
we can show that these equations becomes the first order equations 
\begin{equation}\nonumber
\dot{A}=1-\frac{A^2}{4C^2},\quad \dot{C}=\frac{A}{2C}.
\end{equation}
which is nothing but (\ref{g21}).

\hspace{2mm}

\noindent Remark: ${\bf R}\times Sp(2)/(Sp(1)\times U(1))$, 
${\bf R}\times SU(3)/(U(1)\times U(1))$

\hspace{2mm}

The metric with $G_2$ holonomy (\ref{g22})
could not be obtained from the proposition. 
We will explain this point using the example of $SU(3)/(U(1)\times U(1))$. 
Let the coordinates and the one-forms 
$\sigma_1,\sigma_2,\sigma_4,\cdots,\sigma_7$ 
of  
$SU(3)/(U(1)\times U(1))$
be the restriction of those of 
$SU(3)/U(1)$ to $\psi=0$ in (\ref{cp2form}).
The three-form (\ref{g2form}) corresponding to the metric (\ref{g22}) 
is constructed with the orthonormal basis
\begin{equation}
e_1=A\sigma_2,\quad
e_2=A\sigma_1,\quad 
e_3=dt\quad,
e_4=C\sigma_5,\quad
e_5=C\sigma_4,\quad
e_6=C\sigma_7,\quad
e_7=C\sigma_6,
\end{equation}
which means, if we assume there exist vector fields of the proposition, 
they must be of the form  
\begin{equation}\label{fsigma}
f{{\partial}\over{\partial r}}, \quad a_i(r)f\sigma_i^* \quad(i=1,2,4,5,6,7)
\end{equation}
with local functions $f_i$ on $SU(3)/(U(1)\times U(1))$. 
On the other hand, if we look for a local volume form on this coset
of the form $F\sigma_1\wedge\sigma_2\wedge\sigma_4\wedge\cdots\wedge\sigma_7$ 
and vector fields of the form $f_iX_i (i=1,2,4,5,6,7)$,
the volume-preserving condition is
\begin{eqnarray}\nonumber
d_{X_1}(Ff_1)-(Ff_1)2A_2^3=0,\quad d_{X_2}(Ff_2)+(Ff_2)2A_1^3=0,
\\\nonumber
d_{X_4}(Ff_4)+(Ff_4)(-A_5^3+3A^8_5)=0,\quad
d_{X_5}(Ff_5)+(Ff_5)(A_4^3-3A^8_4)=0,
\\
d_{X_6}(Ff_6)+(Ff_6)(A_7^3+3A^8_7)=0,\hskip 0.2 in
d_{X_7}(Ff_7)+3(Ff_7)(-A_6^3-3A^8_6)=0.
\end{eqnarray}
where $A_i^3$ are the coefficients of $\sigma_i$ in  $A_3$ which is 
the restriction of $\sigma_3$ of $SU(3)/U(1)$ to $\psi=0$.
Using (\ref{identity2}) and the identity 
\begin{equation}
-A_2^3\sigma_1+A_1^3\sigma_2-A_5^3\sigma_4+A_4^3\sigma_5+A_7^3\sigma_6
-A_6^3\sigma_7=
-{{d|w|^2}\over{2(1+|w|^2)}}+{{d|w_1|^2}\over{2(1+|w_1|^2)}}
-{{1}\over{2}}d\log\sin\theta,
\end{equation} the solution is
\begin{eqnarray}
Ff_1&=&\mbox{function of }w,\theta,\\
Ff_2&=&\sin\theta \mbox{ up to multiplication by function of }w_1,w_2,\nonumber\\
Ff_4=Ff_5=Ff_6=Ff_7&=&(1+|w|^2)^{5/4}(1+|w_1|^2)^{-1/2}(\sin\theta)^{1/2}
\mbox{ up to const.}\nonumber
\end{eqnarray}  
Obviously $f_2$ can not be equal to $f_4=\cdots=f_7$.
Hence we could not find the vector fields of the form (\ref{fsigma}) 
and the construction of the proposition can not be 
applied in this case. 
$Sp(2)/(Sp(1)\times U(1))$ case can be explained similarly.
We should note that in general we can consider the ansatz
on ${\bf R} \times SU(3)/(U(1)\times U(1))$
\begin{equation}
ds^2=dt^2+A(t)^2\left(\sigma_1^2+\sigma_2^2\right)
+C_1(t)^2\left(\sigma_4^2+\sigma_5^2\right)
+C_2(t)^2\left(\sigma_6^2+\sigma_7^2\right).
\end{equation}
We have to set $C_1=C_2$ for $Sp(2)/(Sp(1)\times U(1))$.

\section{Conclusions and Discussions}
\hspace{5mm}

In this paper, we have discussed the coset construction
of the metrics with exceptional holonomy
on ${\bf R}\times {\rm coset \; spaces}$
by using the formulation with volume-preserving vector fields. 
For the cosets which give the metrics with $Spin(7)$
holonomy, we have determined explicit volume-preserving vector fields.
We also have discussed the metrics with $G_2$ holonomy in a similar way. 
We wish to know the application of these results.
Purely theoretically,
it would be very interesting to develop the method 
appliciable for the ansatz beyond the coset spaces,
and find the generalization of the formulation with vector 
fields to the $SU(4), Sp(2)$ holonomy in eight-dimensions.

We may expect the other generalization in the metrics with $G_2$ holonomy.
It was argued in \cite{Acharya} that the configuration of
D6-branes wrapping special Lagrangian ${\bf S}^3$
in the deformed conifold $T^*{\bf S}^3$ \cite{Cade} can be represented 
as the M-theory compactification on the space of 
${\bf R}^4$ bundle over ${\bf S}^3$ with $G_2$ holonomy.
For the solution with $Spin(7)$ holonomy \cite{CvGiLuPo},
the metrics with $G_2$ holonomy in \cite{GiPaPo} have been 
naturally appeared.
Then, one may think natural to expect the similar
Taub-NUT and Taub-BOLT generalization of the metric 
on ${\bf R}^4$ bundle on ${\bf S}^3$. 
However, the answer seems negative;
If we adopt the following ansatz for the vector fields
which is the generalization of (\ref{vec2}):
\begin{eqnarray}
&&V_1=c(r)\left(\sigma_1^*+\frac{1}{2}\Sigma_1^*\right),\quad
V_2=c(r)\left(\sigma_2^*+\frac{1}{2}\Sigma_2^*\right),\quad
V_3=b(r)\left(\sigma_3^*+\frac{1}{2}\Sigma_3^*\right),\nonumber\\
&&V_4=a(r)\Sigma_1^*,\quad
V_5=a(r)\Sigma_2^*,\quad
V_6=a(r)\Sigma_3^*, \quad V_7=\frac{\partial}{\partial r},\nonumber
\end{eqnarray}
the resulting equations necessarily reduce to $c=b$.
 
Finally, it is natural to ask how the multi-center metrics
are given by using the vector fields.
In four-dimensions, the multi-center metric \cite{GiHa} can be
derived from the following ansatz for the vector fields 
on ${\bf R}\times {\bf R}^3$ with the volume form
$d\tau \wedge dx^1 \wedge dx^2\wedge dx^3$ \cite{AsJaSm}
\begin{eqnarray}
V_0=V(\vec{x})\frac{\partial}{\partial\tau},\quad
V_i=\frac{\partial}{\partial x_i}
-\omega_i(\vec{x})\frac{\partial}{\partial\tau},\nonumber
\end{eqnarray}
where $i=1,2,3$.
However, the naive generalization to the higher dimensions does not work.
Nevertheless,
there is some existence theorem for multi-center metric with $SU, Sp$ holonomy 
proved by Joyce in \cite{Joyce}.
For this purpose, it seems promising to consider the HyperK\"ahler quotient
construction for metrics with $Sp(2)$ holonomy along the line of 
\cite{HiKaLiRo, Polchinski}.

\section*{Acknowledgements}
\hspace{5mm}
We would like to thank B.S. Acharya, T. Eguchi, T. Ootsuka
and C.N. Pope for useful communications.
Especially, we are grateful to Y. Yasui for the important comments.
The research of M.N. is supported by the JSPS fellowships for Young Scientists. 
\section*{Note Added}
\hspace{5mm}
After submitted this article, many new complete non-singular
solutions have been appeared \cite{CvGiLuPo3,
BrGoGuGu, KaYa, CvGiLuPo4, GuSp}.

For $Spin(7)$ metrics,
the equation $(8)$ in \cite{CvGiLuPo4} has been derived in this paper.
We have presented more general first order equations 
with five radial functions for $SU(3)/U(1)$ case.
We expect by \cite{CvGiLuPo4}
that the equation (3.42) would give the solution
which behaves like Atiyah-Hithin metric in four-dimensions.
In \cite{CvGiLuPo4}, the authors have considered
the Wallach spaces $N_{k,\ell}\simeq SU(3)/U(1)$.   
We have presented the analysis for $k=\ell$ case.
It should be straightforward to do $k\not=\ell$ case.

The generalization on $G_2$ holonomy metric
mentioned in conclusion is pursued in \cite{BrGoGuGu}.
In general, we may consider the ansatz with the different basis
invariant under the change between two $SU(2)$ left-invariant one-forms
\cite{CvGiLuPo3, BrGoGuGu}. 
It is straightforward to derive the first order equations for
this ansatz by the method in section 4.

\section*{A $\;$ Left-invariant One-forms}
\hspace{5mm}

We introduce the left-invariant vector 
fields and the left-invariant 1-forms on 
classical groups and see how they descend to coset spaces.

Let $G$ be $SU(m),Sp(m)$ or $SO(m)$  and 
let $n$ be the real dimension of $G$. 
We denote the orthogonal basis of the 
Lie algebra $Lie G$ of $G$ by 
$T_1,T_2,\cdots,T_n$ with respect to the 
positive definite inner product $<T_i,T_j>$ on $Lie G$ 
We denote the structure constant by $C_{ijk}$ 
;
\begin{equation}
[T_i,T_j]=\sum_{k=1}^n C_{ijk}T_k.
\end{equation}
Let $A_{ij} (1\leq i,j \leq n)$ 
be  complex functions on $G$ which assign 
 an element $a\in G $ its $(i,j)$-th entry, 
i.e., $A_{ij}(a)=a_{ij}$.

Consider the matrix-valued 1-form $A^{-1}dA$ 
called the Cartan canonical 1-form. 
It is $Lie G$-valued and we write it in the following form;
\begin{equation}
A^{-1}dA=\sum_{i=1}^n \widetilde{\sigma}_i T_i.
\end{equation}
Then $\widetilde{\sigma}_1,\cdots,\widetilde{\sigma}_n$
are linearly independent, globally defined 1-forms on 
$G$ and invariant under the left action of $G$ on $G$.
They are called left-invariant 1-forms.
The differentials of $\widetilde{\sigma}_i$'s are
\begin{equation}\label{dtildesigma}
d\widetilde{\sigma}_k=-\sum_{1\leq i<j\leq n}C_{ijk}
\widetilde{\sigma}_i\wedge \widetilde{\sigma}_j.
\end{equation}
Vector fields
dual to $\{\widetilde{\sigma}_1,\cdots,\widetilde{\sigma}_n\}$
are also invariant under the left action of $G$ on $G$ and 
called left-invariant vector fields.
They generate the right action of $G$ on $G$. 

Now we turn to the space of orbits $G/H$ with respect 
to the right action of $H$ on $G$ where $H$ is a $n-\ell$ dimensional 
closed subgroup of $G$. 
If we choose a representative of each $H$-orbit smoothly, 
we can identify it as a point in  some neighborhood $U\subset G/H$.
We shall change the basis of $Lie G$ 
if needed so that $T_{\ell+1},\cdots,T_n$ span the basis of $Lie H$.
Then 
$\widetilde{\sigma}_{\ell+1}^{*},\cdots,\widetilde{\sigma}_n^*$ are 
generators of the right $H$-action.  
Let $\sigma_1,\cdots,\sigma_n$ be the restriction of 
$\widetilde{\sigma}_1,\cdots,\widetilde{\sigma}_n$ .
$\{\sigma_1,\cdots,\sigma_{\ell}\}$ is the basis of $T^*(G/H)$ and
we can write $\sigma_{\ell+1},\cdots,\sigma_{n}$ as the  linear combinations :
\begin{equation}
\sigma_{i}=\sum_{k=1}^{\ell}A^i_k\sigma_k
\hskip 0.3 in \ell+1\leq i\leq n.
\end{equation}
Thus the defferential of $\sigma_1,\cdots,\sigma_{\ell}$ are
\begin{equation}\label{dsigma}
d\sigma_k=-\sum_{1\leq i<j\leq \ell}C_{ijk}\sigma_i\wedge \sigma_j
          -\sum_{1\leq i,j\leq \ell}\left(\sum_{p=\ell+1}^n C_{ipk}A^p_j\right)
          \sigma_i\wedge\sigma_j.
\end{equation}
We choose the basis of the tangen space of $G/H$, 
$\{X_1,\cdots,X_{\ell}\}$ as the dual of $\sigma_1,\cdots,\sigma_{\ell}$;
\begin{equation}\label{sx}
\sigma_i(X_j)=\delta_{i,j}
\hskip 0.3 in 1\leq i,j \leq \ell.
\end{equation} 
Then the Lie brackets of these vector fields becomes
\begin{equation}\label{liebrackets}
[X_i,X_j]=\sum_{k=1}^{\ell}C_{ijk}X_k+
          \sum_{k+1}^{\ell}\left(\sum_{p=\ell+1}^nA^p_jC_{ipk}\right)X_k,
\end{equation}
by the identity 
$[X_i,X_j]+\sum_{k=1}^{\ell}d\sigma_k(X_i,X_j)X_k=0$
which folllows from (\ref{sx}).

In general the 1-forms $\sigma_1,\cdots,\sigma_{\ell}$ 
and the vector fields $X_1,\cdots,X_{\ell}$ 
exist only locally on 
$G/H$. That is because the Cartan 1-form 
is not invariant under the right action of $H$ in general, 
so  when we change the representatives of 
$H$-orbits in $G$, $\sigma_i$'s also change.
Thus we must be careful about the global 
existence of a metric on the orbit space 
when constructing it 
from the restriction of left-invariant 1-forms. 

\section*{B $\;$ Coset Spaces $Sp(2)/Sp(1)$ and $SU(3)/U(1)$}
\hspace{5mm}

{\it $Sp(2)/Sp(1)$}

\hspace{2mm}

First we define the coordinates on $Sp(2)$ 
and compute the left-invariant one-forms $\sigma_i$ 
following the appendix A.
Then we will choose the representative of the 
$Sp(1)$-orbits and obtain one-forms $\sigma_i$ on $Sp(2)/Sp(1)$.

We will present the basis of
the Lie algebra $sp(2)$ we used in the main sections 
and write down an explicit
local trivialization of $Sp(2)$ as the $Sp(1)$
principal bundle over ${\bf S}^4$. 

For simplicity, we use quaternion definition of 
the symplectic group and its Lie algebra $sp(n)$
;
$Sp(n)=\{n\times n \hskip 0.1 in 
\mbox{quaternion matrix such that} 
\hskip 0.1 in A^{\dagger}A=1\}$, 
$sp(n)=\{n\times n \hskip 0.1 in 
\mbox{quaternion matrix such that} 
\hskip 0.1 in T^{\dagger}+T=0 \}$.
We take the following basis for $sp(2)$:
$$
T_1=\left(
\begin{array}{cc}
\bf{i}&0\\0&0
\end{array}
\right),
\hskip 0.2 in
T_2=\left(
\begin{array}{cc}
-\bf{j}&0\\0&0
\end{array}
\right),
\hskip 0.2 in 
T_3=\left(
\begin{array}{cc}
\bf{k}&0\\0&0
\end{array}
\right),
$$$$
T_4={{1}\over{\sqrt{2}}}\left(
\begin{array}{cc}
0&-\bf{i}\\-\bf{i}&0
\end{array}
\right),
\hskip 0.1 in
T_5={{1}\over{\sqrt{2}}}\left(
\begin{array}{cc}
0&\bf{j}\\\bf{j}&0
\end{array}
\right),
\hskip 0.1 in
T_6={{1}\over{\sqrt{2}}}\left(
\begin{array}{cc}
0&\bf{k}\\\bf{k}&0
\end{array}
\right),
\hskip 0.1 in
T_7={{1}\over{\sqrt{2}}}\left(
\begin{array}{cc}
0&-1\\1&0
\end{array}
\right),
$$$$
T_8=\left(
\begin{array}{cc}
0&0\\0&\bf{i}
\end{array}
\right),
\hskip 0.2 in 
T_9=\left(
\begin{array}{cc}
0&0\\0&-\bf{j}
\end{array}
\right),
\hskip 0.2 in 
T_{10}=\left(
\begin{array}{cc}
0&0\\0&\bf{k}
\end{array}
\right).
$$
The commutation relations are
\begin{eqnarray}\nonumber
[T_1,T_2]=-2T_3,\hskip 0.2 in
[T_1,T_3]=2T_2,\hskip 0.2 in
[T_1,T_4]=-T_7,
\\\nonumber
[T_1,T_5]=T_6,\hskip 0.2 in
[T_1,T_6]=-T_5,\hskip 0.2 in
[T_1,T_7]=T_4,
\\\nonumber
[T_1,T_8]=[T_1,T_9]=[T_1,T_{10}]=0,
\\\nonumber
[T_2,T_3]=-2T_1,\hskip 0.2 in 
[T_2,T_4]=-T_6,\hskip 0.2 in 
[T_2,T_5]=-T_7,\hskip 0.2 in
[T_2,T_6]=T_4,\hskip 0.2 in
[T_2,T_7]=T_5,
\\\nonumber
[T_2,T_8]=[T_2,T_9]=[T_2,T_{10}]=0
\\\nonumber
[T_3,T_4]=-T_5,\hskip 0.2 in
[T_3,T_5]=T_4,\hskip 0.2 in
[T_3,T_6]=T_7,\hskip 0.2 in
[T_3,T_7]=-T_6,
\\\nonumber
[T_3,T_8]=[T_3,T_9]=[T_3,T_{10}]=0
\\\nonumber
[T_4,T_5]=-T_3-T_{10},\hskip 0.2 in
[T_4,T_6]=-T_2-T_9,\hskip 0.2 in
[T_4,T_7]=-T_1-T_8,
\\\nonumber
[T_4,T_8]=-T_7,\hskip 0.2 in 
[T_4,T_9]=T_6,\hskip 0.2 in
[T_4,T_{10}]=T_5,
\\\nonumber
[T_5,T_6]=T_1+T_8,\hskip 0.2 in
[T_5,T_7]=-T_2-T_9,\hskip 0.2 in
\\\nonumber
[T_5,T_8]=-T_6,\hskip 0.2 in
[T_5,T_9]=-T_7,\hskip 0.2 in
[T_5,T_{10}]=-T_4,
\\\nonumber
[T_6,T_7]=T_3+T_{10},\hskip 0.2 in
[T_6,T_8]=T_5,\hskip 0.2 in
[T_6,T_9]=-T_4,\hskip 0.2 in 
[T_6,T_{10}]=T_7,
\\\nonumber
[T_7,T_8]=T_4,\hskip 0.2 in 
[T_7,T_9]=T_5,\hskip 0.2 in
[T_7,T_{10}]=-T_6,
\\\nonumber
[T_8,T_9]=-2T_{10},\hskip 0.2 in
[T_8,T_{10}]=2T_9,\hskip 0.2 in 
[T_9,T_{10}]=-2T_8.
\end{eqnarray}
If we see the first column of the $Sp(2)$-matrix 
as a quaternion line in ${\bf H}^2$ or a point in ${\bf S}^4$, 
then coordinates of $Sp(2)$ as the principal 
$Sp(1)\times Sp(1)$-bundle can be chosen as follows 
when $A_{11}\neq 0(A\in Sp(2))$:
\begin{equation}
A=
{{1}\over{\sqrt{1+|x|^2}}}
\left(
\begin{array}{cc}
h_1&-\overline{ x}h_2\\
xh_1&h_2
\end{array}
\right),
\hskip 0.2 in
|h_1|^2=|h_2|^2=1.
\end{equation}
Here, $x
=x_0+x_1{\bf i}+x_2{\bf j}+x_3{\bf k}
$ is 
a quaternion coordinate of ${\bf HP}^1={\bf S}^4$ and 
$h_1,h_2$ are quaternions of norm 1.
We write $h_1$ using the $SU(2)$ coordinates $\theta,\psi,\phi$ as follows;
\begin{equation}
h_1=\cos{{\theta}\over{2}}e^{{{{\bf k}}\over{2}}(\psi+\phi)}
+{\bf j}\sin{{\theta}\over{2}}e^{{{{\bf k}}\over{2}}(\psi-\phi)},
\end{equation}
where $\theta,\psi,\phi$ have period $2\pi, 4\pi, 2\pi$, respectively.

The left-invariant one-forms are
\begin{eqnarray}
{\bf i}\widetilde{\sigma}_1-{\bf j}\widetilde{\sigma}_2+{\bf k}\widetilde{\sigma}_3=
\overline{h}_1dh_1
+{{\overline{h}_1({\overline x} dx-d\overline{x}x )h_1}\over{2(1+|x|^2)}}
,
\nonumber\\
-{\bf i} \widetilde{\sigma}_4+{\bf j}\widetilde{\sigma}_5+{\bf k}\widetilde{\sigma}_6
={{1}\over{\sqrt{2}}}\left({{\overline{h}_2dx h_1}\over{1+|x|^2}}-
{{\overline{h}_1d\overline{x}h}_2\over{1+|x|^2}}\right),
\nonumber\\
\widetilde{\sigma}_7={{1}\over{\sqrt{2}}}\left(
{{\overline{h}_2dx h_1}\over{1+|x|^2}}+
{{\overline{h}_1d\overline{x}h_2}\over{1+|x|^2}}\right),
\nonumber\\
{\bf i}\widetilde{\sigma}_8-{\bf j}\widetilde{\sigma}_9+{\bf k}
\widetilde{\sigma}_{10}=
\overline{h}_2dh_2
+{{\overline{h}_2 (xd\overline{x}-d\overline{x}x) h_2}\over{2(1+|x|^2)}}
.
\end{eqnarray}
Note that 
${\bf i}\tilde{\sigma}_1-{\bf j}\tilde{\sigma}_2+{\bf k}\tilde{\sigma}_3$
consists of  the canonical 1-forms on $Sp(1)=SU(2)$
\begin{equation}
\overline{h}_1dh_1={{{\bf k}}\over{2}}(d\psi+\cos\theta d\phi)
+{{e^{-{\bf k}\phi}}\over{2}}(-d\theta-{\bf k}\sin\theta d\phi)(-{\bf j})
\end{equation}
and the self-dual connection over ${\bf S}^4$
\begin{equation}
{\bf i}{\cal A}_1-{\bf j}{\cal A}_2+{\bf k}{\cal A}_3
={{\overline{x}dx-d\overline{x}x}\over{2(1+|x|^2)}},
\end{equation}
rotated by the adjoint action of $Sp(1)$. 
Here, the orientation of ${\bf S}^4$ is $x_1, x_2, x_3, x_0$.
By (\ref{dtildesigma}),
the differential of the left-invariant one-forms are 
read from the structure constants of the Lie algebra: 
\begin{eqnarray}\label{dif1}
d{\widetilde{\sigma}}_1&=&
2\widetilde{\sigma}_2\wedge\widetilde{\sigma}_3+\widetilde{\sigma}_4\wedge\widetilde{\sigma}_7
-\widetilde{\sigma}_5\wedge\widetilde{\sigma}_6,
\nonumber\\
d\widetilde{\sigma}_2&=&
-2\widetilde{\sigma}_1\wedge\widetilde{\sigma}_3+\widetilde{\sigma}_4\wedge\widetilde{\sigma}_6
+\widetilde{\sigma}_5\wedge\widetilde{\sigma}_7,
\nonumber\\
d\widetilde{\sigma}_3&=&
2\widetilde{\sigma}_1\wedge\widetilde{\sigma}_2+\widetilde{\sigma}_4\wedge\widetilde{\sigma}_5
-\widetilde{\sigma}_6\wedge\widetilde{\sigma}_7,
\nonumber\\
d\widetilde{\sigma}_4&=&
-\widetilde{\sigma}_1\wedge\widetilde{\sigma}_7-\widetilde{\sigma}_2\wedge\widetilde{\sigma}_6
-\widetilde{\sigma}_3\wedge\widetilde{\sigma}_5+\widetilde{\sigma}_5\wedge\widetilde{\sigma}_{10}
+\widetilde{\sigma}_6\wedge\widetilde{\sigma}_9-\widetilde{\sigma}_7\wedge\widetilde{\sigma}_8,
\nonumber\\
d\widetilde{\sigma}_5&=&
\widetilde{\sigma}_1\wedge\widetilde{\sigma}_6+\widetilde{\sigma}_3\wedge\widetilde{\sigma}_4
-\widetilde{\sigma}_2\wedge\widetilde{\sigma}_7-\widetilde{\sigma}_4\wedge\widetilde{\sigma}_{10}
-\widetilde{\sigma}_6\wedge\widetilde{\sigma}_8-\widetilde{\sigma}_7\wedge\widetilde{\sigma}_9,
\nonumber\\
d\widetilde{\sigma}_6&=&
-\widetilde{\sigma}_1\wedge\widetilde{\sigma}_5+\widetilde{\sigma}_2\wedge\widetilde{\sigma}_4
+\widetilde{\sigma}_3\wedge\widetilde{\sigma}_7+\widetilde{\sigma}_5\wedge\widetilde{\sigma}_8
-\widetilde{\sigma}_4\wedge\widetilde{\sigma}_9+\widetilde{\sigma}_7\wedge\widetilde{\sigma}_{10},
\nonumber\\
d\widetilde{\sigma}_7&=&
\widetilde{\sigma}_1\wedge\widetilde{\sigma}_4+\widetilde{\sigma}_2\wedge\widetilde{\sigma}_5
-\widetilde{\sigma}_3\wedge\widetilde{\sigma}_6+\widetilde{\sigma}_4\wedge\widetilde{\sigma}_8
+\widetilde{\sigma}_5\wedge\widetilde{\sigma}_9-\widetilde{\sigma}_6\wedge\widetilde{\sigma}_{10},
\end{eqnarray}

For $Sp(2)/Sp(1)$,
we chose $\widetilde{\sigma}_8^*,\widetilde{\sigma}_9^*,\widetilde{\sigma}_{10}^*$ 
as the generators 
of $Sp(1)$ and $h_2=1$ as the representative of  the
$Sp(1)$-orbits when $A_{11}\neq 0$.
Then 
$\sigma_1,\cdots,\sigma_7$ which are 
the restriction of $\widetilde{\sigma}_1,\cdots,\widetilde{\sigma}_7$
span $T^*(Sp(2)/Sp(1))$. They are
\begin{eqnarray}
{\bf i}\sigma_1-{\bf j}\sigma_2+{\bf k}\sigma_3&=&
\overline{h}_1dh_1
+{{\overline{h}_1({\overline x} dx-d\overline{x})x h_1}\over{2(1+|x|^2)}}
,\nonumber\\
-{\bf i} \sigma_4+{\bf j}\sigma_5+{\bf k}\sigma_6+\sigma_7
&=&\sqrt{2}{{dx h_1}\over{1+|x|^2}}.
\end{eqnarray}
And
\begin{equation}
{\bf i}\sigma_8-{\bf j}\sigma_9+{\bf k}\sigma_{10}
=
{{x d\overline{x}-d{x}\overline{x} }\over{2(1+|x|^2)}}.
\end{equation}
We denote the coefficients of $\sigma_4,\sigma_5,\sigma_6,\sigma_7$
in $\sigma_8,\sigma_9,\sigma_{10}$ by $A^8_4,A^8_5,A_6^8,A_7^8,\cdots,
A_{7}^{10}$. They are
$$
A^8_4={{1}\over{\sqrt{2}}}\Re xh_1,\hskip 0.2 in
A^8_5={{-1}\over{\sqrt{2}}}\Re xh_1{\bf k},\hskip 0.2 in
A^8_6={{1}\over{\sqrt{2}}}\Re xh_1{\bf j},\hskip 0.2 in
A^8_7={{-1}\over{\sqrt{2}}}\Re xh_1{\bf i},
$$$$
A^9_4={{1}\over{\sqrt{2}}}\Re xh_1{\bf k},\hskip 0.2 in
A^9_5={{1}\over{\sqrt{2}}}\Re xh_1,\hskip 0.2 in
A^9_6={{1}\over{\sqrt{2}}}\Re xh_1{\bf i},\hskip 0.2 in
A^9_7={{1}\over{\sqrt{2}}}\Re xh_1{\bf j},
$$\begin{equation}
A^{10}_4={{1}\over{\sqrt{2}}}\Re xh_1{\bf j},\hskip 0.2 in
A^{10}_5={{1}\over{\sqrt{2}}}\Re xh_1{\bf i},\hskip 0.2 in
A^{10}_6={{-1}\over{\sqrt{2}}}\Re xh_1,\hskip 0.2 in
A^{10}_7={{-1}\over{\sqrt{2}}}\Re xh_1{\bf k}.
\end{equation}
We can show the identity (\ref{identity1}) from these coefficients. 

The differential of $\sigma_1,\cdots,\sigma_7$ are of 
the same form as those of 
$\widetilde{\sigma}_1,\cdots,\widetilde{\sigma}_7$, but now,
$\sigma_8,\sigma_9,\sigma_{10}$ are linear combinations of 
$\sigma_1,\cdots,\sigma_7$.
The Lie brakets (\ref{xixjsp}) of the vector fields $X_1,\cdots,X_7$ 
dual to $\sigma_1,\cdots,\sigma_7$ are obtained 
from $A_i^j$, the structure constants and (\ref{liebrackets}).

\hspace{2mm}

{\it $SU(3)/U(1)$}

\hspace{2mm}

Here, we will present the basis of 
the Lie algebra $su(3)$ we used in the main sections 
and write down an explicit
local trivialization of $SU(3)$ as the $SU(2)/{\bf Z}_2$-
principal bundle over ${\bf CP}^2$.

We use the following basis for $su(3)$ :
$$
T_1=\left(
\begin{array}{ccc}
0&i&0\\i&0&0\\0&0&0
\end{array}
\right),\quad
T_2=\left(
\begin{array}{ccc}
0&1&0\\-1&0&0\\0&0&0
\end{array}
\right),\quad
T_3=\left(
\begin{array}{ccc}
i&0&0\\0&-i&0\\0&0&0
\end{array}
\right),
$$
$$
T_4=\left(
\begin{array}{ccc}
0&0&i\\0&0&0\\i&0&0
\end{array}
\right),\;
T_5=\left(
\begin{array}{ccc}
0&0&1\\0&0&0\\-1&0&0
\end{array}
\right),\;
T_6=\left(
\begin{array}{ccc}
0&0&0\\0&0&i\\0&i&0
\end{array}
\right),\;
T_7=\left(
\begin{array}{ccc}
0&0&0\\0&0&1\\0&-1&0
\end{array}
\right),
$$
\begin{equation}
T_8=\left(
\begin{array}{ccc}
i&0&0\\0&i&0\\0&0&-2i
\end{array}
\right).
\end{equation}
The commutation relations are
$$
[T_1,T_2]=-2T_3,\hskip 0.2 in
[T_1,T_3]=2T_2,\hskip 0.2 in
[T_1,T_4]=-T_7,
$$$$
[T_1,T_5]=T_6,\hskip 0.2 in
[T_1,T_6]=-T_5,\hskip 0.2 in
[T_1,T_7]=T_4,
$$$$
[T_1,T_8]=0,
$$$$
[T_2,T_3]=-2T_1,\hskip 0.2 in 
[T_2,T_4]=-T_6,\hskip 0.2 in 
[T_2,T_5]=-T_7,\hskip 0.2 in
[T_2,T_6]=T_4,\hskip 0.2 in
[T_2,T_7]=T_5,
$$$$
[T_2,T_8]=0,
$$$$
[T_3,T_4]=-T_5,\hskip 0.2 in
[T_3,T_5]=T_4,\hskip 0.2 in
[T_3,T_6]=T_7,\hskip 0.2 in
[T_3,T_7]=-T_6,
$$$$
[T_3,T_8]=0,
$$$$
[T_4,T_5]=-T_3+T_{8},\hskip 0.2 in
[T_4,T_6]=-T_2 \hskip 0.2 in
[T_4,T_7]=-T_1,
$$$$
[T_4,T_8]=3T_5,\hskip 0.2 in 
$$$$
[T_5,T_6]=T_1,\hskip 0.2 in
[T_5,T_7]=-T_2,\hskip 0.2 in
[T_5,T_8]=-3T_4,
$$\begin{equation}
[T_6,T_7]=T_3+T_8,\hskip 0.2 in
[T_6,T_8]=3T_7,\hskip 0.2 in
[T_7,T_8]=-3T_6. 
\end{equation}

The following is a coordinate system on $SU(3)$ when 
the (3,3)-th entry is non-zero:
$$
A=\left(
\begin{array}{ccc}
\alpha v_1+\beta v_2,&
-\overline{\beta}v_1+\overline{\alpha} v_2,&
v_3
\end{array}
\right),  
$$$$
v_1={{e^{-it/2}}\over{\sqrt{1+|w_1|^2}}}
\left(\begin{array}{c}
1\\0\\-\overline{w_1}
\end{array}
\right),
$$$$
v_2={{e^{-it/2}}\over{\sqrt{(1+|w_1|^2)(1+|w|^2)}}}
\left(\begin{array}{c}
-\overline{w_2}w_1\\1+|w_1|^2\\-\overline{w_2}
\end{array}
\right),
$$\begin{equation}
v_3={{e^{it}}\over{\sqrt{1+|w|^2}}}
\left(\begin{array}{c}
w_1\\w_2\\1
\end{array}
\right),
\end{equation}
where $|w|^2$ denotes $|w_1|^2+|w_2|^2$.
The third column of $SU(3)$-matrix is 
a vector in ${\bf C}^3$ defining a point in ${\bf CP}^2$ and 
$w_1,w_2$ are  its complex coordinates. 
$t$ is a coordinate of $U(1)$ generated by $T_8$,
 and $\alpha,\beta$ are complex numbers satysfying 
 $|\alpha|^2+|\beta|^2=1$ which can be regarded as coordinates of 
$SU(2)$:when $\alpha\neq 0$,we can  write them in the 
following form;
\begin{equation}
\alpha=\cos{{\theta}\over{2}}e^{i(\psi+\phi)/2},
\hskip 0.2 in
\beta=\sin{{\theta}\over{2}}e^{i(\psi-\phi)/2}.
\end{equation}
Note that because $(\alpha,\beta,t)$ and 
$(-\alpha,-\beta,t+2\pi)$ are a same point,
they are rather coordinates of $SU(2)/{\bf Z}_2=SO(3)$ and 
the periods of $(\theta,\psi,\phi)$ should be  $(2\pi,2\pi,2\pi)$ rather  
than $(2\pi,4\pi,2\pi)$.

The left-invariant one-forms are
\begin{eqnarray}
i\widetilde{\sigma}_1+\widetilde{\sigma}_2&=&
e^{-i\psi}\Biggl[{{1}\over{2}}\left(-d\theta-i\sin\theta d\phi)\right.
\nonumber\\
&&\; + \cos^2{{\theta}\over{2}}e^{-i\phi}(i{\cal A}_1+{\cal A}_2)
+\sin^2{{\theta}\over{2}}e^{-i\phi}(-i{\cal A}_1+{\cal A}_2)
-i\sin\theta{\cal A}_3\Biggr],
\nonumber
\\
\widetilde{\sigma}_3&=&{{1}\over{2}}
(d\psi+\cos\theta d\phi)
+\cos\theta{\cal A}_3
+\sin\theta\;\Im [e^{-i\phi}(i{\cal A}_1+{\cal A}_2)]
\nonumber\\
\left(
\begin{array}{c}
i\widetilde{\sigma}_4+\widetilde{\sigma}_5\\
i\widetilde{\sigma}_6+\widetilde{\sigma}_7
\end{array}
\right)&=&
\left(\begin{array}{cc}
\cos{{\theta}\over{2}}e^{-i(\psi+\phi)/2}&
\sin{{\theta}\over{2}}e^{-i(\psi-\phi)/2}\\
-\sin{{\theta}\over{2}}e^{i(\psi-\phi)/2}&
\cos{{\theta}\over{2}}e^{i(\psi+\phi)/2}
\end{array}
\right)
\left(\begin{array}{c}
{{e^{3it/2}dw_1}\over{\sqrt{(1+|w_1|^2)(1+|w|^2)}}}\\
{{e^{3it/2}(-\overline{w}_1w_2dw_1+(1+|w_1|^2)dw_2)}
\over{(1+|w|^2)\sqrt{1+|w_1|^2}}}
\end{array}
\right)
,\nonumber\\
\tilde{\sigma}_8&=&{{i}\over{4}}
\left({{\overline{w}_1dw_1+\overline{w}_2dw_2}
\over{(1+|w|^2)}}-c.c.\right)
-{{1}\over{2}}dt,
\end{eqnarray}
where
$$
i{\cal A}_1+{\cal A}_2=
{{-\overline{w}_2dw_1}\over{(1+|w_1|^2)\sqrt{1+|w|^2}}}
,$$
\begin{equation}
{\cal A}_3={{1}\over{2i}}\left(
-{{\overline{w}_1d{w}_1}\over{1+|w_1|^2}}+
{{\overline{w}_1dw_1+\overline{w}_2dw_2}\over{2(1+|w|^2)}}-c.c.
\right)
.\end{equation}
${\cal A}_1,{\cal A}_2,{\cal A}_3$ is the self-dual $SO(3)$-instanton 
on ${\bf CP}^2$.

The differential of the left-invariant one-forms are computed 
by using (\ref{dtildesigma}):
\begin{eqnarray}
d\widetilde{\sigma}_1&=&
2\widetilde{\sigma}_2\wedge\widetilde{\sigma}_3+\widetilde{\sigma}_4\wedge\widetilde{\sigma}_7
-\widetilde{\sigma}_5\wedge\widetilde{\sigma}_6,
\nonumber\\
d\widetilde{\sigma}_2&=&
-2\widetilde{\sigma}_1\wedge\widetilde{\sigma}_3+\widetilde{\sigma}_4\wedge\widetilde{\sigma}_6
+\widetilde{\sigma}_5\wedge\widetilde{\sigma}_7,
\nonumber\\
d\widetilde{\sigma}_3&=&
2\widetilde{\sigma}_1\wedge\widetilde{\sigma}_2+\widetilde{\sigma}_4\wedge\widetilde{\sigma}_5
-\widetilde{\sigma}_6\wedge\widetilde{\sigma}_7,
\nonumber\\
d\widetilde{\sigma}_4&=&
-\widetilde{\sigma}_1\wedge\widetilde{\sigma}_7-\widetilde{\sigma}_2\wedge\widetilde{\sigma}_6
-\widetilde{\sigma}_3\wedge\widetilde{\sigma}_5+3\widetilde{\sigma}_5\wedge\widetilde{\sigma}_8,
\nonumber\\
d\widetilde{\sigma}_5&=&
\widetilde{\sigma}_1\wedge\widetilde{\sigma}_6+\widetilde{\sigma}_3\wedge\widetilde{\sigma}_4
-\widetilde{\sigma}_2\wedge\widetilde{\sigma}_7-3\widetilde{\sigma}_4\wedge\widetilde{\sigma}_8,
\nonumber\\
d\widetilde{\sigma}_6&=&
-\widetilde{\sigma}_1\wedge\widetilde{\sigma}_5+\widetilde{\sigma}_2\wedge\widetilde{\sigma}_4
+\widetilde{\sigma}_3\wedge\widetilde{\sigma}_7+3\widetilde{\sigma}_7\wedge\widetilde{\sigma}_8,
\nonumber\\
d\widetilde{\sigma}_7&=&
\widetilde{\sigma}_1\wedge\widetilde{\sigma}_4+\widetilde{\sigma}_2\wedge\widetilde{\sigma}_5
-\widetilde{\sigma}_3\wedge\widetilde{\sigma}_6-3\widetilde{\sigma}_6\wedge\widetilde{\sigma}_8.
\end{eqnarray}
 
For $SU(3)/U(1)$,
we choose $\widetilde{\sigma}^*_8$ as the generators of $U(1)$ action 
and choose t=0 as the representative of the $U(1)$-orbits.
Then 
$\sigma_1,\cdots,\sigma_7$ which is the restriction of 
$\tilde{\sigma}_1,\cdots,\tilde{\sigma}_7$ is the basis 
of $T^*(SU(3)/U(1))$. They are
\begin{eqnarray}
i\sigma_1+\sigma_2&=&e^{-\psi}\Biggl[
{{1}\over{2}}(-d\theta-i\sin\theta d\phi)\nonumber\\
&&\;+ i\cos\theta\;\Im[e^{-i\phi}(i{\cal A}_1+{\cal A}_2)]
+\sin\theta\;\Re[e^{-i\phi}(i{\cal A}_1+{\cal A}_2)]
-i\sin\theta {\cal A}_3\Biggr],
\nonumber\\
{\sigma}_3&=&{{1}\over{2}}
(d\psi\cos\theta d\phi)
+\cos\theta{\cal A}_3
-\Im [\sin\theta e^{i\phi}(-i{\cal A}_1+{\cal A}_2)],
\nonumber\\
\left(
\begin{array}{c}
i\sigma_4+\sigma_5\\
i\sigma_6+\sigma_7
\end{array}
\right)&=&
\left(\begin{array}{cc}
\cos{{\theta}\over{2}}e^{-i(\psi+\phi)/2}&
\sin{{\theta}\over{2}}e^{-i(\psi-\phi)/2}\\
-\sin{{\theta}\over{2}}e^{i\psi-\phi)/2}&
\cos{{\theta}\over{2}}e^{i(\psi+\phi)/2}
\end{array}
\right)
\left(\begin{array}{c}
{{dw_1}\over{\sqrt{(1+|w_1|^2)(1+|w|^2)}}}\\
{{(-\overline{w}_1w_2dw_1+(1+|w_1|^2)dw_2)}
\over{(1+|w|^2)\sqrt{1+|w_1|^2}}}
\end{array}
\right).
\end{eqnarray}
And 
\begin{equation}
\sigma_8={{i}\over{4}}
\left({{\overline{w}_1dw_1+\overline{w}_2dw_2}
\over{2(1+|w|^2)}}-c.c.\right).
\end{equation}
can be written as the linear combination of others:
\begin{eqnarray}
\sigma_8&=&{{-1}\over{2\sqrt{1+|w_1|^2}}}
\left[
\Re\left(
\overline{w_2}\sin{{\theta}\over{2}}e^{-i\phi/2}
+\sqrt{1+|w|^2}\overline{w_1}\cos{{\theta}\over{2}}e^{i\phi/2}
\right)\sigma_4\right.\nonumber\\
&+&\Im\left(
\overline{w_2}\sin{{\theta}\over{2}}e^{-i\phi/2}
+\sqrt{1+|w|^2}\overline{w_1}\cos{{\theta}\over{2}}e^{i\phi/2}
\right)\sigma_5
\nonumber\\
&+&\Re\left(
\overline{w_2}\cos{{\theta}\over{2}}e^{-i\phi/2}
+\sqrt{1+|w|^2}\overline{w_1}\sin{{\theta}\over{2}}e^{i\phi/2}
\right)\sigma_6\nonumber\\
&+&\left.\Im\left(
\overline{w_2}\cos{{\theta}\over{2}}e^{-i\phi/2}
+\sqrt{1+|w|^2}\overline{w_1}\sin{{\theta}\over{2}}e^{i\phi/2}
\right)\sigma_7
\right]
\end{eqnarray}
We denote the coefficients of $\sigma_4, \dots, \sigma_7$ 
by $A_4^8,\cdots,A_7^8$.
We can show that the identity (\ref{identity2}) holds.

The differential of $\sigma_i$'s (\ref{dsigma2}) are of the same form 
as $\widetilde{\sigma}_i$'s.
The Lie brackets of the vector fields $X_1, \cdots, X_7$ (\ref{xixjsu}) 
which are dual of $\sigma_1, \cdots,\sigma_7$ 
are computed by using (\ref{liebrackets})

\hspace{2mm}

$SU(2)$

\hspace{2mm}

The basis of the Lie algebra $su(2)$ is
\begin{equation}
T_1=\left(\begin{array}{cc}
0&i\\i&0
\end{array}
\right),\quad
T_2=\left(\begin{array}{cc}
0&1\\-1&0
\end{array}
\right),\quad
T_3=\left(\begin{array}{cc}
i&0\\0&-i
\end{array}
\right).
\end{equation}
The differential of the three left-invariant one forms 
$\widetilde{\sigma}_1,\widetilde{\sigma}_2,\widetilde{\sigma}_3$ are
\begin{equation}
d\widetilde{\sigma}_1=2\widetilde{\sigma}_2\wedge\widetilde{\sigma}_3,\quad
d\widetilde{\sigma}_2=-2\widetilde{\sigma}_1\wedge\widetilde{\sigma}_3,\quad
d\widetilde{\sigma}_3=2\widetilde{\sigma}_1\wedge\widetilde{\sigma}_2.
\end{equation}
We can write $A\in SU(2)$ as 
\begin{equation}
A=\left(\begin{array}{cc}\alpha&-\overline{\beta}\\
\beta&\overline{\alpha}\end{array}
\right),\quad 
\alpha=\cos{{\theta}\over{2}}e^{i(\psi+\phi)/2},\quad
\beta=\sin{{\theta}\over{2}}e^{i(\psi-\phi)/2},
\end{equation}
where $(\theta,\psi,\phi)$ have the periods $(2\pi,4\pi,2\pi)$. 
Then the left-invariant one-forms are
\begin{equation}
i\sigma_1+\sigma_2={{e^{-i\psi}}\over{2}}
(-d\theta-i\sin\theta d\phi),\hskip 0.2 in 
\sigma_3={{1}\over{2}}(d\psi+\cos\theta d\phi).
\end{equation}

\end{document}